\documentclass[ superscriptaddress, aps, prb, twocolumn, floatfix]{revtex4-2}

\usepackage[ usenames, dvipsnames ]{xcolor}
\usepackage{amsmath, amsfonts, amssymb, amsthm, amstext, bm, bbm}
\usepackage[colorlinks=true, linkcolor=blue,citecolor=red, urlcolor=blue, hypertexnames=true]{hyperref}
\usepackage{graphicx}
\usepackage{caption}                                        
\usepackage{physics}
\usepackage{ragged2e}
\DeclareCaptionJustification{justified}{\justifying}
\usepackage[normalem]{ulem}

\newcommand{\p}{\rm p}
\newcommand{\brac}[1]{\left( #1 \right) }

\newcommand{\LL}{{\mathcal{L}}}
\newcommand{\LO}{{\mathcal{L}_0}}
\newcommand{\LJ}{{\mathcal{L}_J}}
\newcommand{\Nm}{{\mathcal{N}_-}}
\newcommand{\N}{{\hat{N}}}

\newcommand{\OO}{{\mathcal{O}}}

\DeclareUnicodeCharacter{2009}{\,}

\begin{document}

\title{Ergodic and chaotic properties in Tavis-Cummings dimer: quantum and classical limit}

\author{Tamoghna Ray}
\email[ ]{tamoghna.ray@icts.res.in }
\affiliation{International Centre for Theoretical Sciences, Tata Institute of Fundamental Research,
Bangalore 560089, India}

% \author{David A. Huse}
% \email[ ]{huse@princeton.edu}
% \affiliation{Physics Department, Princeton University, Princeton, NJ, 08544, USA}

\author{Manas Kulkarni}
\email[ ]{manas.kulkarni@icts.res.in}
\affiliation{International Centre for Theoretical Sciences, Tata Institute of Fundamental Research,
Bangalore 560089, India}

\date{\today}

\begin{abstract}
We investigate two key aspects of quantum systems by using the Tavis-Cummings dimer system as a platform. The first aspect involves unraveling the relationship between the phenomenon of self-trapping (or lack thereof) and integrability (or quantum chaos). Secondly, we uncover {the possibility of} mixed behavior in this quantum system using diagnostics based on random matrix theory and make an in-depth study of classical-quantum correspondence. The setup chosen for the study is precisely suited as it (i) enables a transition from delocalized to self-trapped states and (ii) has a well-defined classical limit, thereby amenable to studies involving classical-quantum conjectures. The obtained classical model in itself has rich chaotic and ergodic properties which were probed via maximal Lyapunov exponents. Furthermore, we present aspects of chaos in the corresponding open quantum system and make connections with non-Hermitian random matrix theory.

% This article focuses on two key aspects of quantum dynamics. Firstly, it explores the relationship between delocalization and quantum chaos, presenting numerical results that show self-trapping in regions where the system exhibits Poisson statistics. Secondly, the study uncovers mixed behavior in the quantum system which is indicated by spectral statistics based on random matrix theory, and investigates its correspondence to the classical limit. The investigation employs the Tavis Cummings dimer system, enabling the transition from a delocalized to a self-trapped state and from chaotic to integrable states within a feasible range of interaction strength. Exact diagonalization is utilized to study the quantum dynamics and evaluate spectral statistics. Tools from random matrix theory, such as the distribution of adjacent gap ratios, level spacing statistics, and spectral form factor, are employed to analyze quantum chaos. Additionally, the well-established method of maximal Lyapunov exponent is utilized to investigate chaos and ergodicity in the classical limit. By addressing these critical aspects, this paper sheds light on intriguing phenomena in the quantum domain and establishes their connections to classical dynamics. Furthermore, the analysis is extended to open quantum systems and an effective non-hermitian model.

\end{abstract}

\maketitle
\section{Introduction}
The study of self-trapping (localization) - delocalization physics in both quantum and classical systems \cite{smerzi1997quantum,albiez2005direct,sarchi2008coherent,venumadhav2010finite,schmidt2010nonequilibrium,raftery2014observation,PhysRevLett.113.174101,secli2021signatures,pouthier2022quantum} is an area of active research. This phenomenon of self-trapping becomes even more interesting when the quantum system under consideration has a well-defined classical limit \cite{polkovnikov2002nonequilibrium,trujillo2009nonequilibrium,pudlik2013dynamics,dey2020engineering}  which provides an excellent platform to investigate potential quantum-classical correspondence. Such a correspondence lies within the realm of quantum chaos and relies on two conjectures, namely the Berry-Tabor conjecture \cite{berry1977level} and the Bohigas-Giannoni-Schmit (BGS) conjecture \cite{PhysRevLett.52.1}, which form the foundation for defining and characterizing quantum chaos. The Berry-Tabor conjecture states that a generic system that is classically integrable and has two or more degrees of freedom has quantum levels that cluster together and the level spacing of adjacent levels has Possoin distribution. On the other hand, the BGS conjecture proposes that classically chaotic systems have quantum levels that are correlated and exhibit repulsion. The spectrum of such systems follows random matrix theory (RMT) statistics.

The set of diagnostic tools to explore the aspects mentioned above stems from RMT \cite{mehta2004random, haake1991quantum,forrester2010log}. More precisely, quantities of interest are level spacing statistics, adjacent gap ratio \cite{oganesyan2007localization}, and the spectral form factor \cite{PhysRevResearch.3.L012019,Cotler_SFFChaos2017,ShenkerGharibyan2018onsetofRM,Liu_SFFChaos_PhysRevD.98.086026}. {There have been experimental developments in measuring such quantities \cite{PhysRevLett.105.024101,assmann2016quantum,frisch2014quantum,doi:10.1126/science.aao1401}. Recently, in Ref.~\onlinecite{dong2024measuring}, the spectral form factor has been experimentally measured using a superconducting quantum processor to probe the presence of (or lack thereof) quantum chaos in quantum many-body systems.} These quantities probe the nature of eigenvalues and the correlations between them. Chaos in many quantum models has been studied extensively \cite{PhysRevLett.122.024101,PhysRevX.12.011018,L_Corps_2022,PhysRevB.105.165432,Nakerst:2022prc,Kirkova_2023}. Several studies have shown a transition from integrability to quantum chaos in different models \cite{PhysRevE.67.066203,PhysRevE.81.036206,WANG2022105222,Prasad_2024}. Other studies include probing symmetries in quantum systems using RMT \cite{PhysRevX.12.011006}. More recently, important progress has been made in characterizing chaos in non-Hermitian Hamiltonians \cite{hamazaki2019non,ghosh2022spectral,PhysRevX.12.021040,PhysRevB.106.064208,PhysRevResearch.4.043196,gupta2023quantum}, open quantum systems \cite{10.1063/5.0082046,Alvaro_2022,hamazaki2022lindbladian,gupta2023quantum}, and connecting to non-Hermitian RMT \cite{PhysRevLett.123.254101,PhysRevX.10.021019}. Here, the quantities of interest include important generalizations of the Hermitian analogs such as the complex spacing ratio \cite{PhysRevX.10.021019} and the dissipative spectral form factor \cite{PhysRevLett.127.170602,PhysRevLett.130.140403}. {Furthermore, with the realization of non-Hermitian systems in experiments \cite{lee2014heralded,lapp2019engineering}, the study of such models becomes even more imperative.}

In the field of {self-trapping-}delocalization physics, the phenomenon of macroscopic self-trapping in a quantum network is of interest for potential {relevance in} quantum computation \cite{wright2019benchmarking, song2019generation, xu2022metrological, lu2019global, figgatt2019parallel, debnath2016demonstration, senko2014coherent} and simulations \cite{xu2020probing,senko2014coherent}. {This phenomenon has been reported in several theoretical as well as experimental studies}. These include bosonic Josephson junctions (BJJs) consisting of cold-atomic Bose-Einstein condensates (BECs) \cite{milburn1997quantum,raghavan1999coherent, smerzi1997quantum,levy2007ac, albiez2005direct,xhani2020dynamical, o2012quantum} and photonic systems \cite{shelykh2008josephson, makin2009time, abbarchi2013macroscopic, coto2015self, schmidt2010nonequilibrium, raftery2014observation,ray2022localization}.
% \sout{Several studies have explored the effects of interactions, connectivity, drive and dissipation, and network size on macroscopic self-trapping showing the transition from a delocalized phase to a localized phase.} 
Circuit and cavity QED models like Jaynes Cummings, Tavis Cummings, Dicke, and Bose Hubbard models serve as interesting platforms to study the effect of non-trivial interactions on localization physics. Several of these models of hybrid quantum systems have a well-defined classical limit making them amenable to exploring classical-quantum conjectures.
% \sout{Even in systems with no interactions, interesting long-range connectivity can lead to nontrivial transitions to self-trapped states which can be related to the physics of flat bands \cite{ray2022localization}.}
Several studies have verified the validity of the BGS and the Berry-Tabor conjecture for various quantum systems. {However, }most of the existing literature has focused on investigating the classical limit when the quantum system exhibits either RMT or Poisson behavior. {Thus there is a need to understand these conjectures when the systems are neither fully RMT nor Poisson class. This is especially important given recent interests in systems that exhibit mixed phases \cite{Nakerst:2022prc,wang2023power,wang2023mixed}.}
% Recently there has been a growing interest in probing quantum systems that exhibit mixed classical phase space . 

This article focuses on two main key {aspects, using the Tavis-Cummings dimer (TCD) model as the platform}. Firstly, it delves into exploring the connection between delocalization and quantum chaos. {In particular, we highlight} instances of self-trapping {(localization)} in {the parameter space} where the system demonstrates Poisson statistics and delocalization when the system exhibits GOE statistics. The second aspect involves the identification of possible mixed behavior in the quantum system, as revealed by {diagnostic tools based on} RMT {where the quantities of interest turn out to be neither GOE nor Poisson. The platform used for this study has a well-defined classical limit. In this work, we make an in-depth study of the resulting classical dimer model and the classical-quantum correspondence.} {For investigating the quantum dynamics (for example, imbalance of excitations) and assessing the spectral statistics, we apply exact diagonalization (ED). Utilizing tools from RMT, including the distribution of adjacent gap ratio, level spacing statistics, and spectral form factor, enables the examination of quantum chaos. Additionally, employing the well-known method of maximal Lyapunov exponent allows for the exploration of chaos and ergodicity in the classical limit.}

This paper is structured as follows. In Sec.~\ref{sec:tcd} we introduce the system of Tavis Cummings dimer in detail. In Sec.~\ref{sec:tot_imbalance} we introduce the total imbalance operator which quantifies the excitation imbalance in the system. We present numerical results showing a transition from a delocalized to a self-trapped state with increasing light-matter interaction. In Sec.~\ref{sec:quantum_chaos} we evaluate the energy spectrum of the system using exact diagonalization and calculate {the level spacing statistics, the distribution of adjacent gap ratio, and the spectral form factor (SFF)}. We present results showing a transition from GOE statistics to Poisson statistics. {We also unravel parameter regimes where the statistics are neither GOE nor Poisson, thereby indicating {the possibility of} mixed behavior.} In Sec.~\ref{sec:classical_limit} we explore the classical limit of the system where we compute the maximal Lyapunov exponent to probe chaos and ergodicity in the system. Finally, we extend our analysis of chaos to open quantum systems. In Sec.~\ref{sec:openTCD} we connect our system to Markovian baths, {mimicking gain-loss dynamics,} using the Lindblad formalism and study the {complex} spectral statistics of the Liouvillian. In Sec.~\ref{sec:NHTCD}, we consider a {non-Hermitian gain-loss Hamiltonian model and also compute its complex spectral statistics. In both the Liouvillian and the non-Hermitian case, we show deep connections to non-Hermitian RMT and 2D Poisson statistics.} Certain details are relegated to the appendices. 

% Motivated by the rich and interesting physics that has been explored in the field of localization-delocalization phenomenon, we investigate its relation to quantum chaos and integrability. Furthermore, in light of the BGS and Berry-Tabor conjecture, we investigate the quantum-classical correspondence within the framework of localization physics. For the purposes of our investigation, we use the Tavis-Cummings dimer (TCD) model which allows a transition from a delocalized to a self-trapped state and also from chaotic to integrable state, within a reasonable range of interaction strength. Using exact diagonalization (ED) we study the dynamics as well as the spectral statistics of the system.

\section{Tavis-Cummings Dimer system}
\label{sec:tcd}
The Tavis Cummings (TC) Hamiltonian models a system of $N$ identical two-level systems (TLS) interacting with a single mode radiation field in the dipole approximation \cite{PhysRev.170.379,kirton2019introduction,10.3389/fphy.2022.980167}. The Hamiltonian ($\hbar = 1$) is given by

\begin{equation}
    H_{\rm TC} = \omega_c a^\dagger a + \omega_s \sum_{i=1}^{N}s^z_i + \frac{\lambda}{\sqrt{N}}\left(a^\dagger \sum_{i=1}^{N}s^-_i + a \sum_{i=1}^{N}s^+_i \right),
\end{equation}
where $\omega_c$ is the frequency of the radiation field in the cavity, $\omega_s$ is the energy gap between the two levels of the TLS and $\lambda$ is the light-matter interaction strength. $a^\dagger$ and $a$ are the cavity creation and annihilation operators, $s^{x,y,z}_i$ are the angular momentum operators for the $i^{\rm th}$ TLS and $s^{\pm}_i = s^x_i \pm i s^y_i$. We define the collective spin operators as,
    $S^z = \sum_{i=1}^{N}s^z_i$, and $S^{\pm} = \sum_{i=1}^{N}s^{\pm}_i$,
which further simplifies the Hamiltonian to
\begin{equation}
    H_{\rm TC} = \omega_c a^\dagger a + \omega_s S^z + \frac{\lambda}{\sqrt{N}}\left(a^\dagger S^- + a S^+ \right).
    \label{eq:TC_hamil}
\end{equation}
We consider the case where the quantum number for $\hat{S}^2$ is set to $S = \sum_{i=1}^N(1/2) = N/2$, i.e., the subspace which is totally symmetric under the exchange of any two spins. As evident from equation Eq.~\eqref{eq:TC_hamil}, the Hamiltonian does not mix different $S$ sectors \cite{PhysRevE.67.066203}.

\begin{figure}[h]
    \centering
    \includegraphics[scale = 0.18]{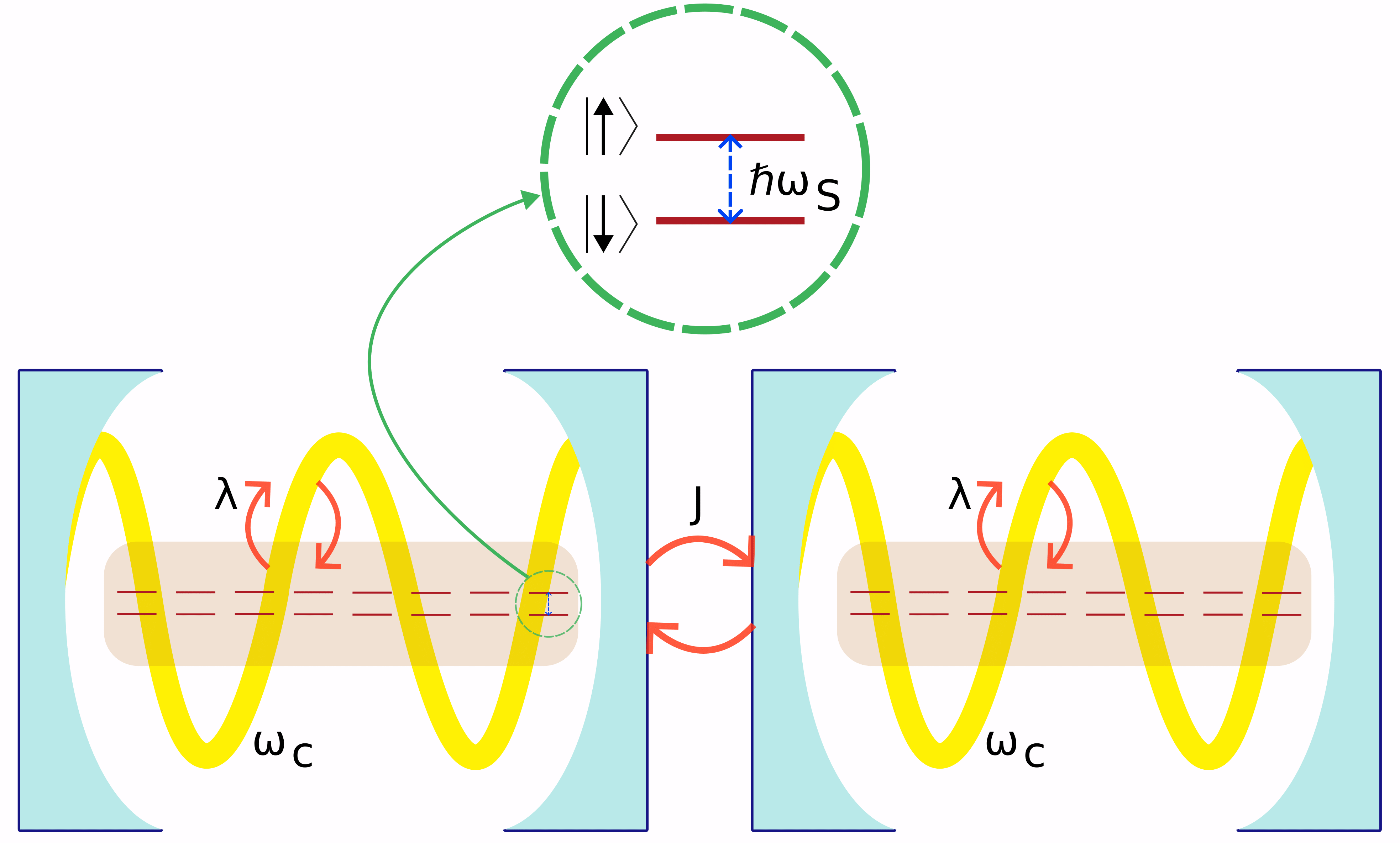}
    \caption{A schematic representation of the Tavis Cummings dimer. The two {units} in the dimer couple to each other with coupling strength $J$. Both the {units} in the dimer are identical with cavity frequency $\omega_c$ and qubit level spacing $\omega_s$. There are $N$ identical atoms {(TLS)} that do not interact with one another {and lead to a large spin system within each unit}. This total spin mode couples to bosonic cavity mode via the interaction $\lambda$. {The Hamiltonian for this setup is given in Eq.~\eqref{eq:TC_dimer}}.}
    \label{TC_scheme}
\end{figure}

In our study, we consider a TC dimer (TCD) [Fig.~\ref{TC_scheme}]. The different TC {units/monomers} (left and right) couple to one another through the cavity modes with strength $J$. There is no direct coupling among the spin modes. The Hamiltonian for such a dimer is given by
\begin{equation}
    H_{\rm TCD} = H_{\rm TC}^L + H_{\rm TC}^R + J\left(a^\dagger_L a_R + a_R^\dagger a_L\right),
    \label{eq:TC_dimer}
\end{equation} 
where $H_{\rm TC}^{L/R}$ is given by Eq.~\eqref{eq:TC_hamil}. {The labels $L$ and $R$ indicate the left and right units respectively. The dimer in Eq.~\eqref{eq:TC_dimer} conserves the total excitation number
\begin{equation}
\hat{N} = a^\dagger_L a_L + S^+_L S^-_L + a^\dagger_R a_R + S^+_R S^-_R\, ,
\label{eq:quantum_excitation}
\end{equation}
whose expectation value is fixed to $N_{\rm p}$. It is worth noting that albeit the single TC unit is integrable \cite{PhysRev.170.379}, the dimer system is non integrable and allows for interesting regimes as one tune the parameter space, in particular $\lambda/J$.}

\section{Imbalance}
\label{sec:tot_imbalance}
{In this section, we introduce the notion of imbalance of excitations.} {In the context of hybrid circuit-QED systems,} localization-delocalization transition has been studied in systems such as the Jaynes-Cummings dimer both theoretically \cite{schmidt2010nonequilibrium} and experimentally \cite{raftery2014observation}. We find that the TCD, which is a {higher spin} generalization of the Jaynes-Cummings dimer, also exhibits a transition from a delocalized state to a self-trapped state by varying $\lambda$ (in units of J). To study this behavior of the excitations, we define the imbalance operator as 
\begin{equation}
    \hat{{I}} = \frac{1}{N_{\rm p}}\brac{a^\dagger_L a_L + S^+_L S^-_L - a^\dagger_R a_R - S^+_R S^-_R},
    \label{eq:total_imbalance}
\end{equation}
which gives the excitation population difference between the left and right monomers, normalized by the {total excitation number $N_{\p}$}. The imbalance at a time $t$ is given by 
\begin{equation}
    {I}(t) = \expval{\hat{{I}}}{\psi(t)}, \text{where}\ket{\psi(t)} = e^{-i H_{\rm TCD}t}\ket{\psi(0)}.
    \label{eq:imbalance_expectation}
\end{equation}
{As mentioned before,} we work in {a fixed} excitation sector $N_{\rm p}$ and initiate the system in an imbalanced condition where all the excitations are localized in the cavity mode of the left monomer, i.e., $\expval{a^\dagger_L a_L} = N_{\rm p}$ and $\expval{S^+_L S^-_L} = \expval{a^\dagger_R a_R} = \expval{S^+_R S^-_R} = 0$.

\begin{figure}
    \centering
    \includegraphics[scale= 0.45]{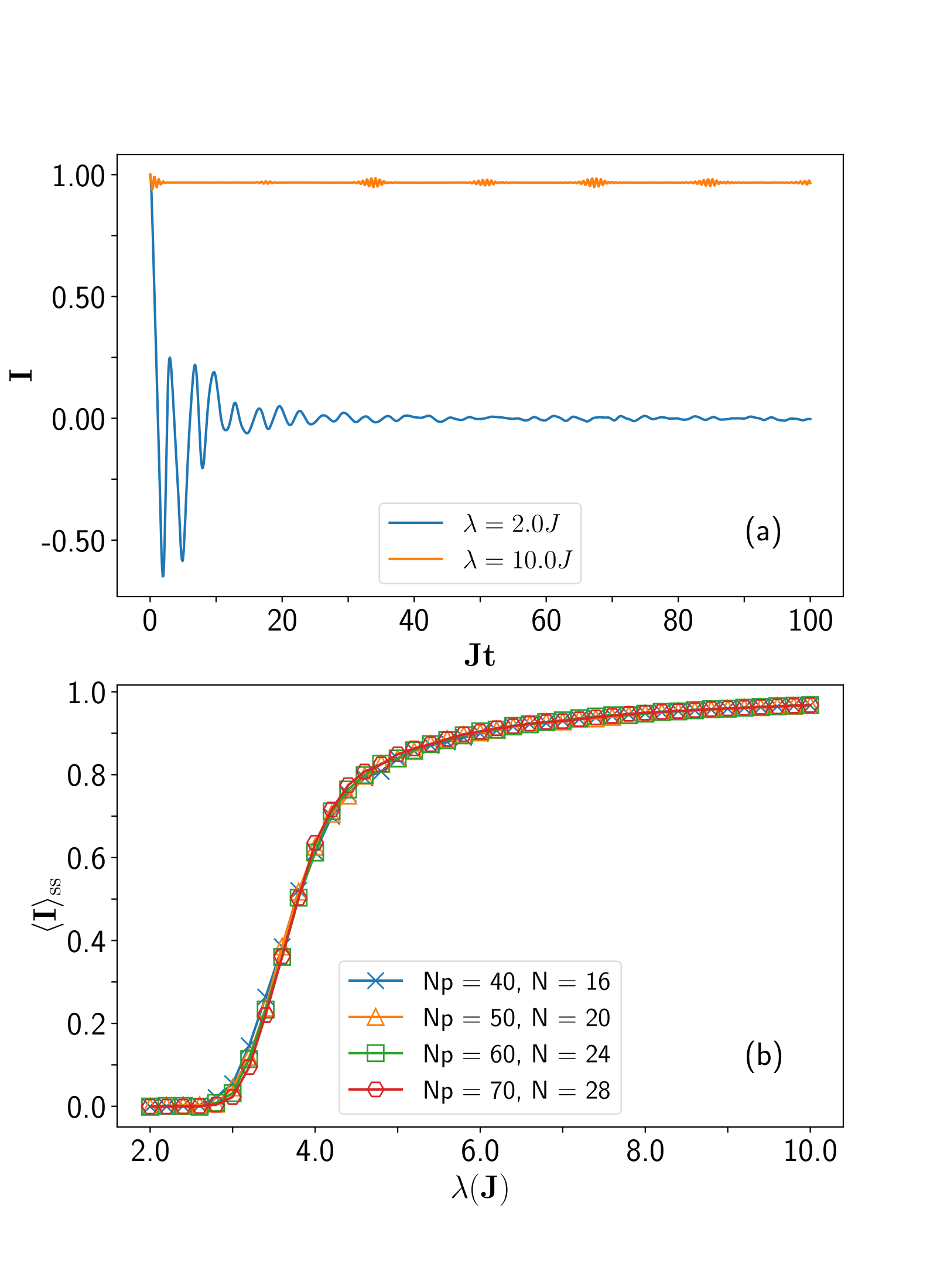}
    \caption{(a) Imbalance for the TCD {defined in Eq.~\eqref{eq:total_imbalance},} with $N_{\p} = 50$ and $N = 20$ is plotted as a function of time {(in units $1/J$)} for $\lambda = 2.0J$ and $10.0J$. (b) Total steady-state imbalance is plotted as a function of $\lambda$ for different system sizes keeping the ratio $N_{\p}/N = 5/2$ fixed. For all the cases we have considered a $10\%$ disorder in the cavity frequency.}
    \label{fig:tot_imbalance}
\end{figure}

In Fig.~\ref{fig:tot_imbalance} (a), {using exact numerics}, we plot ${I}$ given in Eq.~\eqref{eq:imbalance_expectation} as a function of $Jt$ for $\lambda = 2.0 J$ and $\lambda = 10.0J$. In the former case, we observe that the system is delocalized with ${I}$ saturating close to zero, and in the latter case, it is self-trapped with ${I}$ saturating close to $N_{\rm p}$. For intermediate values of $\lambda$, we find that the imbalance saturates to values between zero and $N_{\rm p}$. As a measure of localization, we evaluate the steady state time average of ${I}$ [Eq.~\eqref{eq:imbalance_expectation}] {which is denoted by $\expval{I}_{\rm SS}$}. In Fig.~\ref{fig:tot_imbalance} (b) we plot $\expval{{I}}_{\rm SS}$ as function of $\lambda$ for various excitations, keeping $N_{\rm p}/N = 5/2$ fixed. We observe a transition from delocalized to localized state with increasing $\lambda$. We notice that all the curves collapse on each other when the ratio $N_{\rm p}/N$ is kept fixed which indicates that the critical point of transition $\lambda_{\rm c}$ is a function of $N_{\rm p}/N$. {Interestingly, from our computations we find that the critical point of transition ($\lambda_c$) from delocalized to self-trapped state has a square root dependence on the ratio $N_{\rm}/N$. We estimate this critical point to be 
\begin{equation}
    \lambda_c \approx 1.9J\sqrt{\frac{N_{\rm p}}{N}},
    \label{eq:lam_crititcal}
\end{equation}
when both $N_{\rm p}$ and $N$ are large enough.}
% \textcolor{blue}{In the {appendix or SM} we show how the transition point $\lambda_{\rm c}$ exactly depends on $N_{\rm p}/S$.}

\begin{figure}
    \centering
    \includegraphics[scale = 0.5]{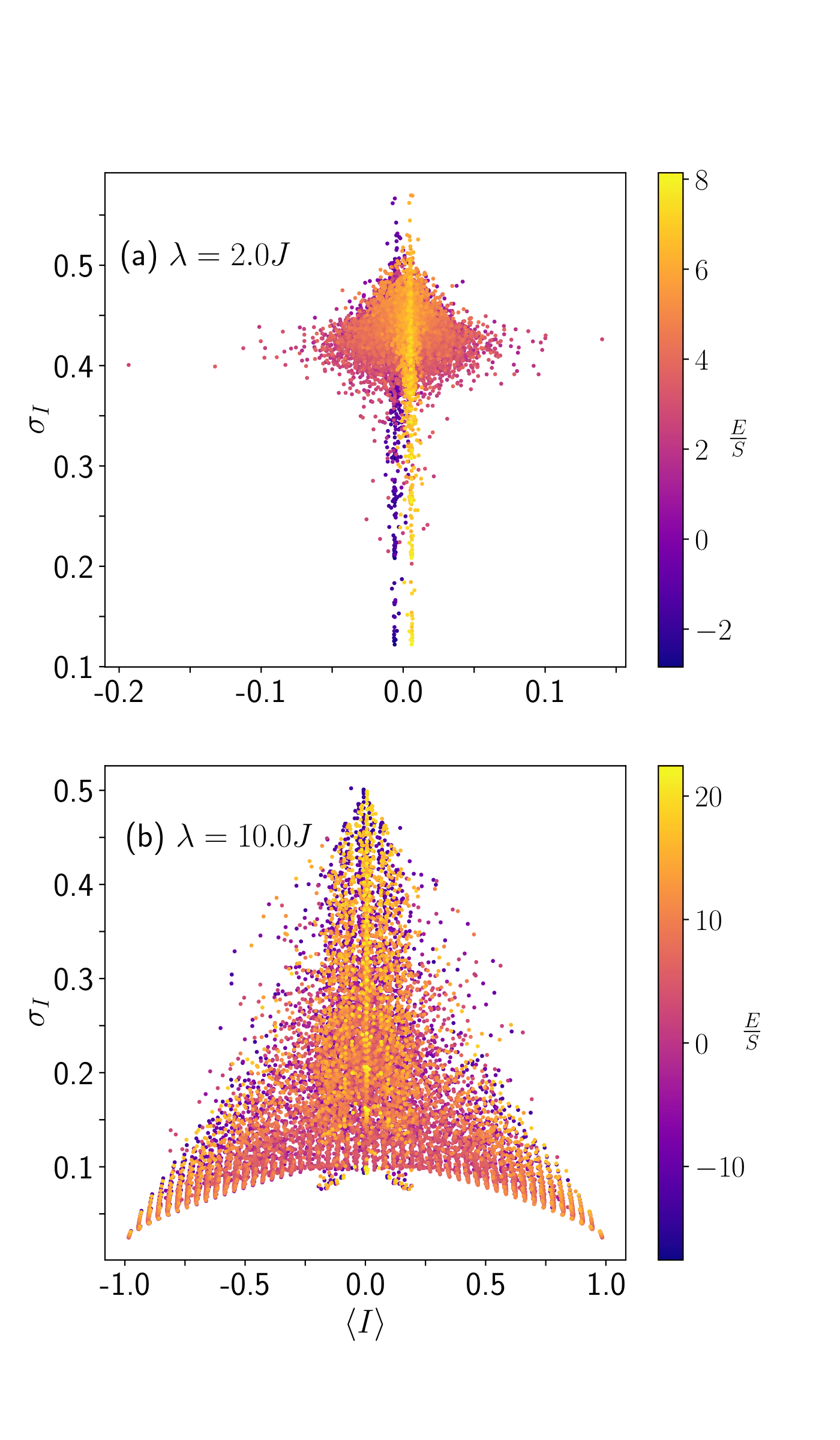}
    \caption{Quantum mechanical standard deviation {[Eq.~\eqref{eq:QM_STD}]} in imbalance ${I}$, $\sigma_{I}$ is plotted as a function of the mean $\expval{{I}}$ for all the eigenvectors of the TC dimer Hamiltonian for (a) $\lambda = 2.0 J$ and (b) $\lambda = 10.0 J$. We take $N_{\rm p} = 50$ and $N = 20$. The energy scaled by $N$ is represented by the color map. {Note the big difference in the x-axis scale between $\lambda=2.0J$ and $\lambda=10.0J$.}}
    \label{fig:quantum mean_vs_std}
\end{figure}

To differentiate between the two cases {(self-trapped and delocalized)}, we further evaluate the imbalance and its standard deviation when the system is in the eigenstates of $H_{\rm TCD}$ [Eq.~\eqref{eq:TC_dimer}]. The quantum mechanical standard deviation for any observable $\mathcal{\hat{O}}$ with respect to the state $\ket{\psi}$ is defined as, 
\begin{equation}
    \sigma_{\hat{\mathcal{O}}} = \sqrt{\expval{\mathcal{\hat{O}}^2}{\psi} - \expval{\mathcal{\hat{O}}}{\psi}^2}.
    \label{eq:QM_STD}
\end{equation} 
In Fig.~\ref{fig:quantum mean_vs_std}, we plot $\sigma_{{I}}$ as a function of $\expval{{I}}$ for all the eigenstates of the TCD for $N_{\rm p} = 50$ and $N = 20$ for (a) $\lambda = 2.0 J$ and (b) $\lambda = 10.0 J$. We consider a $10\%$ disorder in the cavity frequency.  We see a stark difference between the two cases. For the $\lambda = 2.0J$ case, the mean imbalance is mostly spread within $\pm 0.05$ with a few scattered points near -0.2 and +0.1. Whereas for the $\lambda = 10.0J$ case, the mean imbalance ranges between $\pm 1$ which {is consistent} with the results obtained in Fig.~\ref{fig:tot_imbalance}. Furthermore, yet another difference between the two cases is the value of $\sigma_{I}$. For $\lambda = 2.0J$, most of the bulk states have a high standard deviation, but for $\lambda = 10.0J$ this is not the case. 

\section{Chaos and integrability}
\label{sec:quantum_chaos}
In this section, we present a detailed analysis of the spectrum $\{E_n \}$ of the TCD and its relation to random matrix theory (RMT). To generate the distribution, we consider 500 realizations of the TCD Hamiltonian with a $10\%$ disorder in the cavity frequency $\omega_c$ {centered around 1. That is, the disorder was introduced by choosing the cavity frequencies from  $\mathcal{U}(0.9,1.1)$, where $\mathcal{U}(a,b)$ is a uniform distribution between $a$ and $b$.} Introducing the disorder ensures that any symmetry such as the mirror symmetry {(i.e., swapping left and right units)} is broken and we do not end up with spurious statistics. The Hamiltonian is constructed in the Fock basis for the given excitation sector $N_{\rm p}$ and the {sorted} eigenspectrum $\{E_n\}$ is evaluated using exact diagonalization. Furthermore, while evaluating the spectral statistics, to capture the typical behavior of the system, we have discarded the eigenvalues on the edges and considered only the middle third of the spectrum. Several diagnostic tools are used to classify the system as either quantum chaotic or integrable. The term ``integrable" here is used in the sense of spectral properties, where the levels show Poisson statistics and should not be misinterpreted for exact solvability or conventional notions of quantum integrability \cite{sutherland2004beautiful,OLSHANETSKY1983313}.  To probe the short-range correlations of the spectrum we compute the distribution of adjacent level spacing ($s_n = E_{n+1} - E_n$) and the adjacent gap ratio \cite{oganesyan2007localization,atas2013distribution,atas2013joint,giraud2022probing}. {To investigate the long-range correlations in the spectrum, we compute the spectral form factor (SFF).}

\begin{figure}[h]
    \centering
    \includegraphics[scale = 0.45]{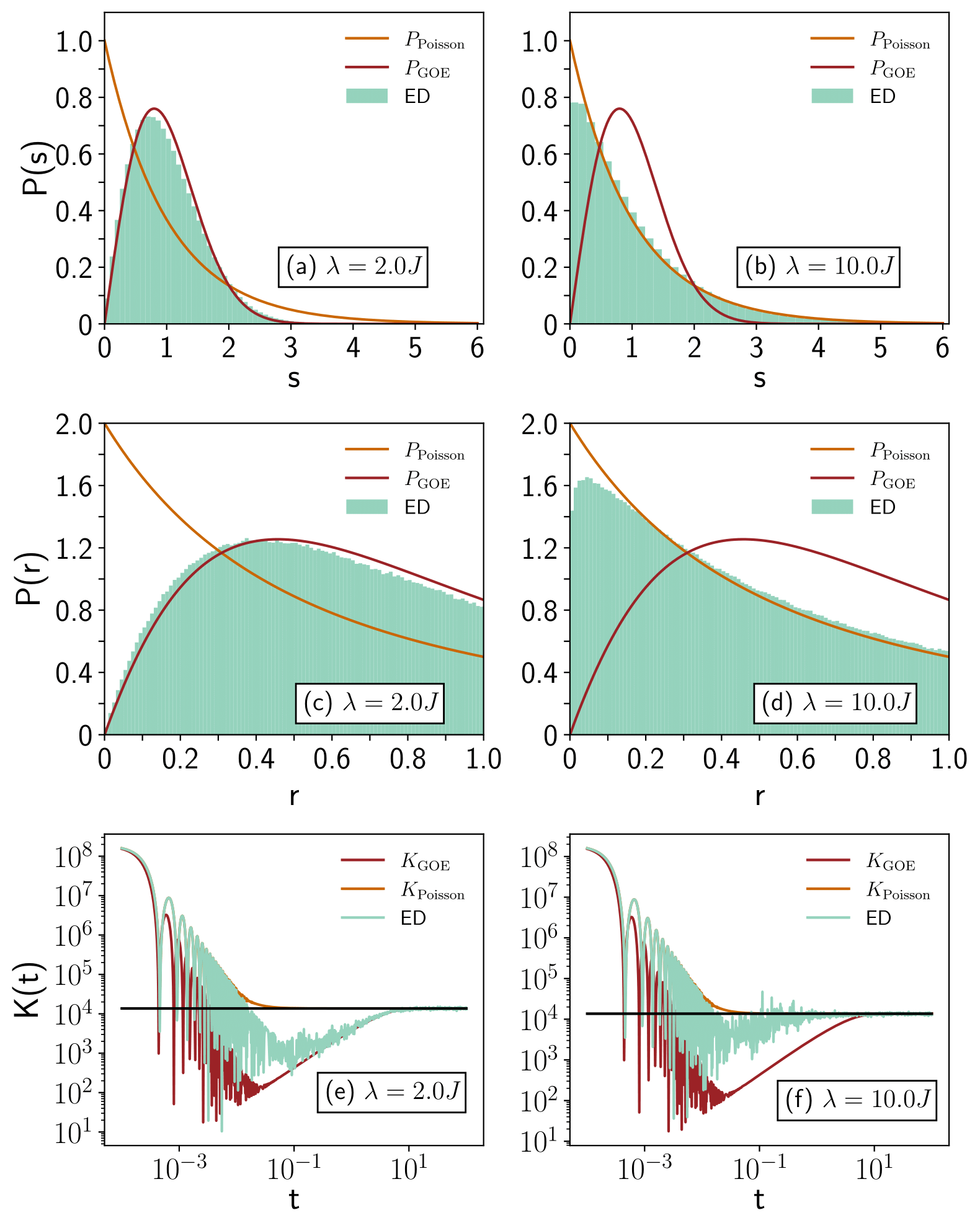}
    \caption{{The spectral analysis described in Sec.~\ref{sec:quantum_chaos} is presented. The} distribution of the adjacent level spacing is plotted for (a) $\lambda = 2.0J$ and (b) $\lambda = 10.0J$ for a TCD [Eq.~\eqref{eq:TC_dimer}]. In (c) and (d), the distribution of the adjacent gap ratio {[Eq.~\eqref{eq:gap_ratio}]} is plotted for $\lambda = 2.0J$ and $\lambda = 10.0J$ respectively. In (e) and (f), the spectral form factor {described in Eq.~\eqref{eq:SFF_eq_main}} is plotted as a function of time (in units of J) for $\lambda = 2.0J$ and $\lambda = 10.0J$ respectively. For all six cases, we have fixed $N_{\rm p} = 50$, $N = 20$ and considered 500 realizations of the $H_{\rm TCD}$ with a $10\%$ disorder in the cavity frequency $\omega_c$.}
    \label{fig:spectral_statistics}
\end{figure}

The adjacent level spacing distribution $P(s)$ is plotted in Fig.~\ref{fig:spectral_statistics}, for (a) $\lambda = 2.0 J$ which agrees with $P_{\rm GOE}(s)$ and (b) $\lambda = 10.0J$ which agrees with $P_{\rm Poisson}(s)$. The theoretical expressions for the two cases are given by
\begin{equation}
    P_{\rm GOE}(s) = \frac{\pi s}{2} e^{-\frac{\pi s^2}{4}},
    \label{eq:level_spacing_GOE}
\end{equation}
\begin{equation}
    P_{\rm Poisson}(s) = e^{-s}.
    \label{eq:level_spacing_Poisson}
\end{equation}
The process of evaluating the level spacing statistics involves unfolding of the spectrum (details in App.~\ref{App:level_spacing}) which can be complicated and often becomes unreliable. To circumvent this issue, Ref.~\onlinecite{oganesyan2007localization} introduced the diagnostic tool of adjacent gap ratio defined as
\begin{equation}
    r_n = \frac{\text{min}\{s_n, s_{n-1}\}}{\text{max}\{s_n, s_{n-1}\}}.
    \label{eq:gap_ratio}
\end{equation}
In Fig.~\ref{fig:spectral_statistics} (c) and (d) we present the distribution of $r$ for both $\lambda  =2.0J$ and $\lambda = 10.0J$. The former agrees with $P_{\rm GOE}(r)$,
\begin{equation}
     P_{\rm GOE}(r) = \frac{27}{4} \frac{r + r^2}{(1 + r + r^2)^{5/2}} \Theta(1-r),
    \label{eq:GOE_gap_ratio}
\end{equation}
and the later agrees with $P_{\rm Poisson}(r)$,
\begin{equation}
    P_{\rm Poisson}(r) = \frac{2}{(1 + r)^2}\Theta(1-r).
    \label{eq:Poisson_gap_ratio}
\end{equation}
Furthermore, we compute the SFF,
\begin{equation}
    K(t,\mathcal{N}) = \expval{\sum_{m,n = 1}^\mathcal{N} e^{i t (E_m - E_n)}},
    \label{eq:SFF_eq_main}
\end{equation}
for both the delocalized and self-trapped regime in Fig.~\ref{fig:spectral_statistics} (e) $\lambda = 2.0J$ and (f) $\lambda = 10.0J$ respectively. {The symbol $\expval{}$ in Eq.~\eqref{eq:SFF_eq_main} denotes average over disorder realizations.} The theoretically predicted behavior of the SFF for the GOE \cite{haake1991quantum,Cotler_SFFChaos2017,ShenkerGharibyan2018onsetofRM,Liu_SFFChaos_PhysRevD.98.086026} and Poisson levels is given by \cite{riser2020nonperturbative, PhysRevResearch.3.L012019,Mahaveer_Abhishodh}
\begin{equation}
    K_{\rm GOE}(t,\mathcal{N}) = K_{\rm GOE}^c(t,\mathcal{N}) + \left[\frac{\pi}{t} J_1\left( \frac{2\mathcal{N}t}{\pi}\right) \right]^2,
\end{equation}
where,
\begin{align}
		\label{SFF_Chaos}
		K_{\rm GOE}^c(t,\mathcal{N} ) &= \mathcal{N}\begin{cases}
			\frac{\mu \tau}{\pi} -\frac{\mu \tau}{2\pi} \log \left(1+ \frac{\mu \tau}{\pi}\right) ~~ 0 <\mu \tau < 2 \pi \\
			2-  \frac{t}{2 \pi} \log \left( \frac{t + \pi}{t -\pi}\right) ~~~2 \pi<t < \infty 
		\end{cases},
\end{align}
and 
\begin{align}
		K_{\rm Poisson}(t,\mathcal{N}) &= \mathcal{N} + \frac{2}{t^2}  - \frac{ (1+i t)^{1-\mathcal{N}} + (1-i t)^{1-\mathcal{N}}  }{t^2}, \label{SFF_Poisson} 
\end{align}
where $J_1(x)$ is the Bessel function of the first kind and $\mathcal{N}$ is the number of eigenvalues considered. As discussed in App.~\ref{app:SFF}, the signatures of chaos indicated by the presence of the ramp in $K(t,\mathcal{N})$ is present in the case of $\lambda = 2.0J$ and is absent in the case of $\lambda = 10.0J$.

\begin{figure}[h]
    \centering
    \includegraphics[scale = 0.45]{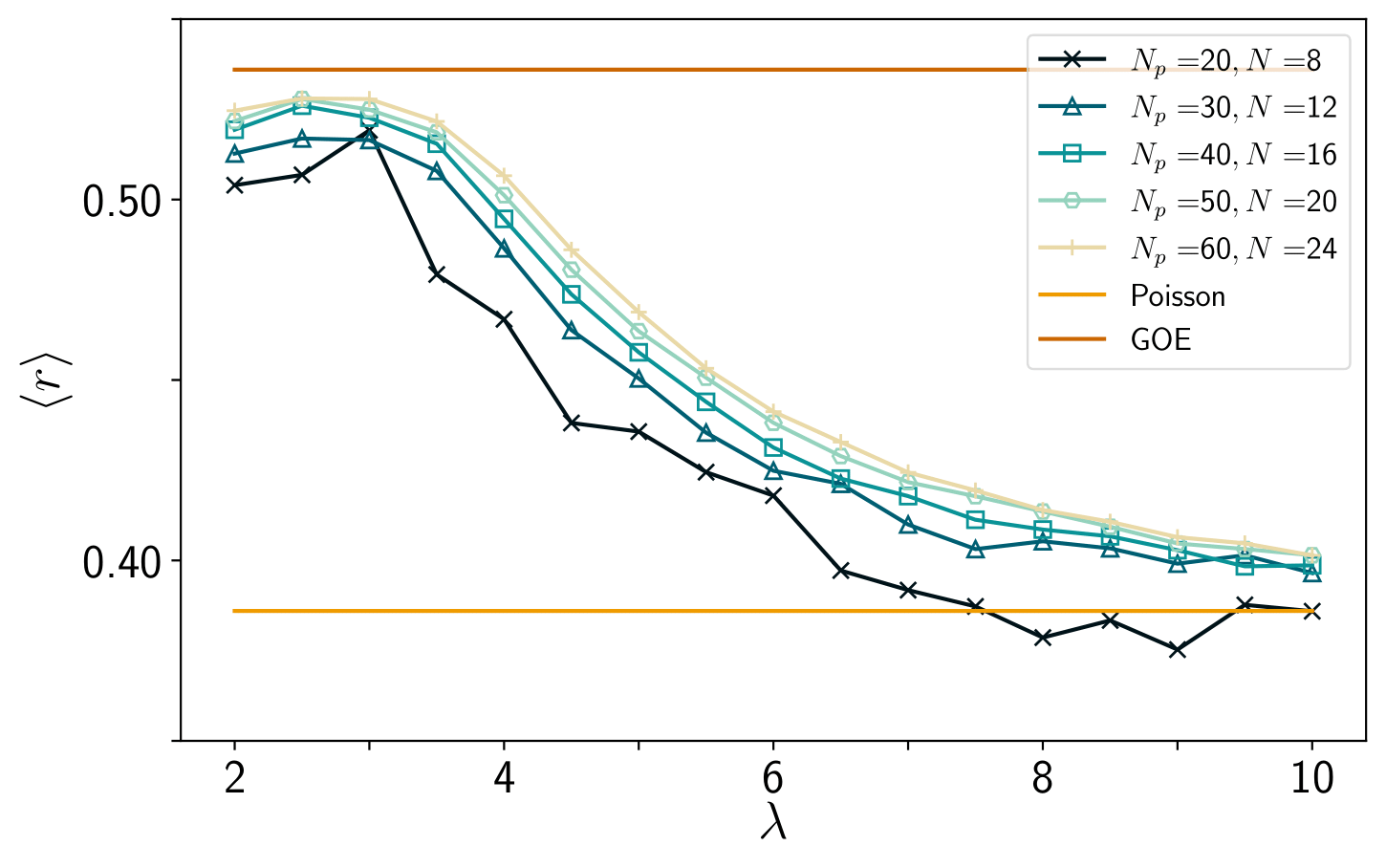}
    \caption{The average adjacent gap ratio $\expval{r}$ {[Eq.~\eqref{eq:gap_ratio}]} is plotted as function of $\lambda$ for $N_{\rm} = 20,30,40,50,$ and 60, keeping the ratio $N_{\rm p}/N = 5/2$ fixed. {We notice that as we increase $N_{\rm p}$ and $N$, keeping their ratio fixed, the curve seems to converge to a limiting shape, which indicates the possible presence of a mixed behavior.} We have generated 500 realizations of $H_{\rm TCD}$ with a $10\%$ disorder in $\omega_c$ for each set of parameters.}
    \label{fig:avg_r_vs_lambda}
\end{figure}

To quantify the chaotic behavior in the system we evaluate the average adjacent gap ratio $\expval{r}$.
For a chaotic system, the average adjacent gap ratio is given by  $\expval{r}_{\rm GOE} \approx 0.536$ and for an integrable system it is given by $\expval{r}_{\rm Poisson} \approx 0.386$. In Fig.~\ref{fig:avg_r_vs_lambda} we plot $\expval{r}$ as function of $\lambda$ for different values of $N_{\rm p}$ keeping $N_{\rm p}/N$ fixed at 5/2. For $\lambda = 0$ the spin modes decouple from the oscillator modes and we get an exactly solvable coupled oscillator model. The spectral statistics for such a limit is likely to be spurious and hence we do not consider $\lambda < 2.0J$. We notice a transition from $\expval{r}_{\rm GOE}$ to $\expval{r}_{\rm Poisson}$ with increasing $\lambda$. Another point of interest is the fact that the transition is not a sharp one and for an intermediate value of $\lambda$ we get neither GOE nor Poisson statistics. This transition does not seem to become sharper for larger local Hilbert space dimensions. {The absence of a sharp transition, for Hilbert space dimensions feasible for computation, indicates that there might be a mixed behavior for intermediate $\lambda$ even in the limit of infinite local Hilbert space. Therefore a naturally interesting question is what happens in the classical limit of the TCD. In the next section [Sec.~\ref{sec:classical_limit}], we show that such mixed behavior is very apparent in the classical limit which is indicated by the coexistence of both regular and chaotic trajectories.}
% An interesting avenue is to investigate the energy dependence of the chaotic behavior \cite{Nakerst:2022prc}.
% In Ref.~\cite{Nakerst:2022prc}, the authors investigate 

\section{Classical limit}
\label{sec:classical_limit}
\begin{figure}
    \centering
    \includegraphics[scale = 0.45]{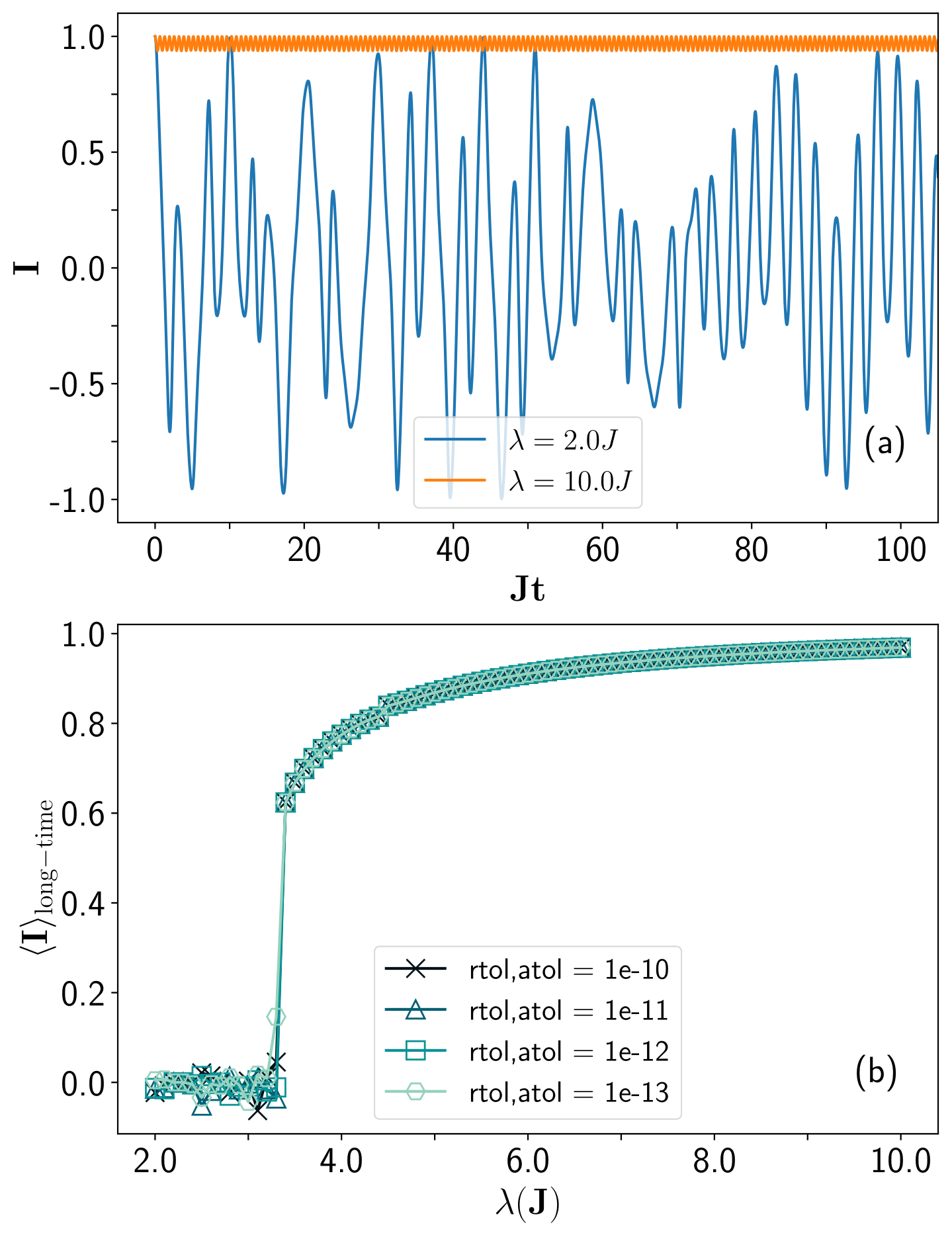}
    \caption{(a) Total classical imbalance for the TCD {defined in Eq.~\eqref{eq:cl_tot_imbalance}} with $N_{\p} = 2.5$ is plotted as a function of time for $\lambda = 2.0J$ and $10.0J$. (b) Total steady-state imbalance is plotted as a function of $\lambda$ for different trajectory precisions. In the legend, ``rtol" and ``atol" stand for relative and absolute tolerances. For all the cases we have considered a $10\%$ disorder in the cavity frequency.}
    \label{fig:classical_tot_imbalance}
\end{figure}

Motivated by the onset of chaos observed in the quantum TCD for small non-zero $\lambda$, {the possible presence of a mixed behavior regime}, and the potential for direct comparison of equivalent quantities, we explore the classical limit of the system. {Such a study is of paramount importance} in the light of the Bohigas, Giannoni, and Schmit (BGS) conjecture. The BGS conjecture proposes that the spectrum of quantized classically chaotic systems follows RMT statistics. {In the case of TCD,} unlike several quantum systems where the classical {limit} is not quite clear, there is a {systematic protocol to go} to the classical limit {from the quantum model. This is detailed in App.~\ref{app:quantum_to_classical}}. The classical TC Hamiltonian is given by,
\begin{equation}
    H_{\rm clTC} = \frac{1}{2}\left(p^2 + \omega_c^2 x^2\right) + \omega_s S^z + \lambda\sqrt{\frac{2}{\omega_c}} \left(\omega_c x S^x - p S^y \right),
    \label{eq:TC_classical_hamil}
\end{equation}
and the TC dimer Hamiltonian is given by,
\begin{equation}
    H_{\rm clTCD} = H_{\rm clTC}^L + H_{\rm clTC}^R + \frac{J}{\omega_c}\left(\omega_c^2 x_Lx_R + p_Lp_R \right),
    \label{eq:TCD_classical}
\end{equation}
{where $H_{\rm clTC}^{L/R}$ is given by Eq.~\eqref{eq:TC_classical_hamil}. The labels $L$ and $R$ indicate the left and right units respectively.} The variables {transform} as 
\begin{eqnarray}
    (x_{L/R},p_{L/R}) &\mapsto& \frac{(x_{L/R}, p_{L/R})}{\sqrt N}\\
    (S^x_{L/R},S^y_{L/R},S^z_{L/R}) &\mapsto& \frac{(S^x_{L/R},S^y_{L/R},S^z_{L/R})}{N},
\end{eqnarray}
 and the energy and imbalance {transform} as
\begin{equation}
    E \mapsto \frac{E}{N}, \hspace{1cm} I \mapsto \frac{I}{N}.
\end{equation}
Therefore, one must scale the values appropriately to compare quantities such as energy or imbalance (which has the same scaling as energy) with the quantum case. We use Eq.~\eqref{eq:TCD_classical} to evaluate the classical equations of motion (EOMs) and numerically solve them to {compute} the total imbalance defined as 
\begin{equation}
    {I}(t) = \frac{\mathcal{N}_L(t) - \mathcal{N}_R(t)}{N_{\rm p}},
    \label{eq:cl_tot_imbalance}
\end{equation}
where the excitation in the left/right monomer is given by
\begin{equation}
    \mathcal{N}_{L/R} = \frac{\omega_c}{2} x_{L/R}^2 + \frac{p_{L/R}^2}{2 \omega_c} + S^z_{L/R} + \frac{1}{2}. 
    \label{eq:excitation}
\end{equation}
We employ the explicit Runge-Kutta method of order 8 \cite{press2007numerical} to solve the classical equations of motion (EOMs) {[see classical EOMs in Eq.~\eqref{eq:classical_EOM}] in App.~\ref{app:quantum_to_classical}}. This algorithm dynamically adjusts the time step to preserve the trajectory {within specified relative and absolute tolerances.} For all classical data, unless otherwise specified, we set both the relative and absolute tolerances to $10^{-12}$. In Fig.~\ref{fig:classical_tot_imbalance} (a), we depict the behavior of ${I}$ against $Jt$ for $\lambda = 2.0J$ and $\lambda = 10.0J$. Here, we observe self-trapping at $\lambda = 10.0J$ and delocalization at $\lambda = 2.0J$, consistent with the quantum case shown in  Fig.~\ref{fig:tot_imbalance}. In Fig.~\ref{fig:classical_tot_imbalance} (b), we illustrate the average long-time imbalance as a function of the interaction strength $\lambda$ for various trajectory precisions. To compute the long-time average of the imbalance, we consider values from $Jt = 0.8\times 10^4$ to $Jt = 10^4$. Once again, we observe good agreement with the quantum case depicted in Fig.~\ref{fig:tot_imbalance}, although the transition point is a bit higher and the transition itself is more abrupt in the classical case. It is noteworthy that for $\lambda$ values lower than the transition point, the agreement among data for different absolute and relative tolerances deteriorates compared to $\lambda$ values higher than the transition point. This outcome aligns with expectations, as in the delocalized regime, numerical errors can magnify due to chaotic system behavior.

\begin{figure}[h]
    \centering
    \includegraphics[scale = 0.38]{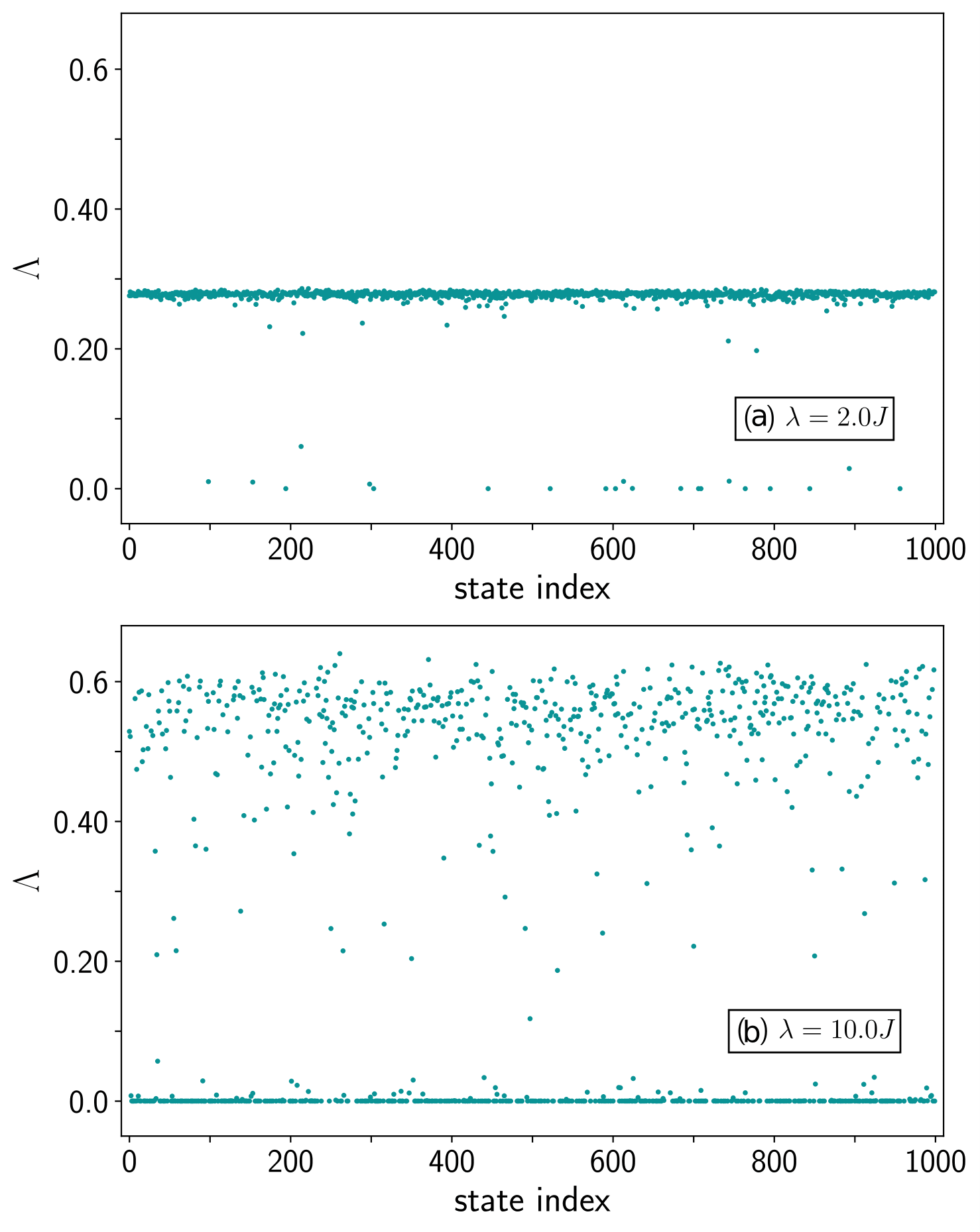}
    \caption{Long time Lyapunov exponent $\Lambda$ is plotted for 500 different initial states for $N_{\p} = 2.5$ and $E = 1.5J$. Two different values of the $\lambda$ are chosen to show the transition from near ergodic to nonergodic phase. }
    \label{fig:lyap_scatter}
\end{figure}

To characterize the chaotic nature of the classical system we compute the Lyapunov exponent $\Lambda$ using the method described in App.~\ref{app:lyap}. To do so, we generate 500 random initial states using the method described in App.~\ref{app:sampling} for each $\lambda$ for a given energy up to a tolerance of $10^{-5}$. In Fig.~\ref{fig:lyap_scatter}, we plot the distribution of the long-time Lyapunov exponent for (a) $\lambda = 2.0J$ and (b) $\lambda = 10.0J$. {For both the values of $\lambda$, an indication of the lack of ergodicity is evident from the scattered nature of the Lyapunov exponents. However, for low values of $\lambda$, the scattering is relatively less, suggesting a more ergodic character in that parameter regime. For $\lambda = 2.0J$, we notice that most of the trajectories are chaotic with a few initial states that lead to regular motion (characterized by non-positive $\Lambda$). Whereas, for $\lambda = 10.0J$, a much larger percentage of the trajectories are regular.}

\begin{figure}[h]
    \centering
    \includegraphics[scale = 0.5]{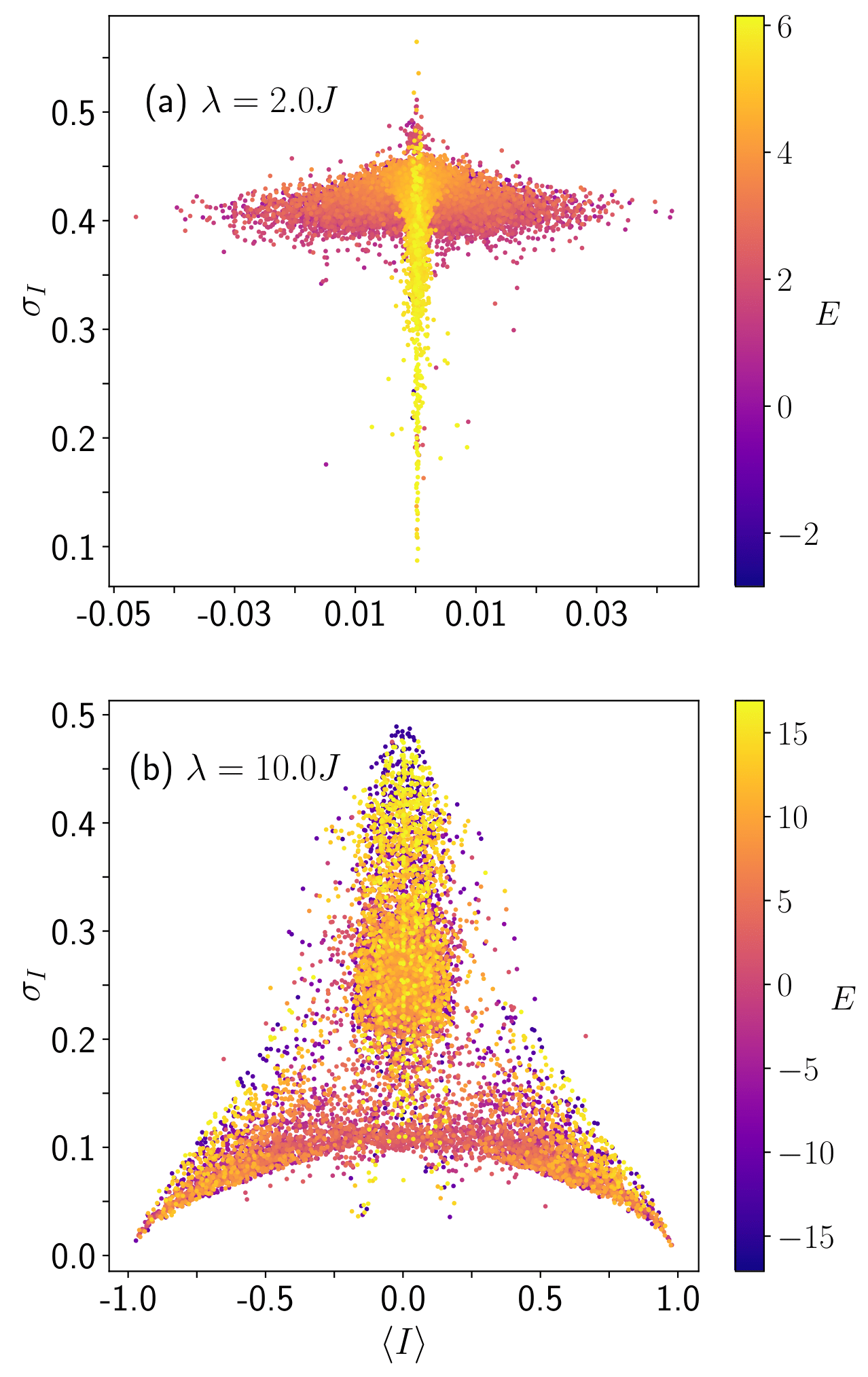}
    \caption{Classical standard deviation in total imbalance ${I}$, $\sigma_{I}$ [Eq.~\eqref{eq:classical_STD}] is plotted as a function of $\expval{{I}}$ for initial states with the same $N_{\p}$ and energy as the eigenstates in Fig.~\ref{fig:quantum mean_vs_std} for (a) $\lambda = 2.0 J$ and (b) $\lambda = 10.0 J$. The energy of the states is represented by the color map. Note the big difference in the x-axis scale between $\lambda = 2.0J$ and $\lambda = 10.0J$.}
    \label{fig:classical_mean_vs_rms}
\end{figure}

{We recall that} in the quantum model (Sec.~\ref{sec:quantum_chaos}), we investigate the imbalance when the system is in the eigenstate of $H_{\rm TCD}$. In the classical model, we do not have a notion of eigenstates. {Motivated to find an analog to Fig.~\ref{fig:quantum mean_vs_std} (quantum case),} we generate multiple random initial states using the method described in App.~\ref{app:sampling}. We choose initial states with the same energy as the eigenstates of $H_{\rm TCD}$ up to a tolerance of $10^{-5}$ and that has a fixed excitation given by $N_{\rm p} = 5/2$. These states are scattered randomly throughout the manifold $N_{\rm p} = 5/2$ in the phase space and are similar to the eigenstates of $H_{\rm TCD}$ which span the fixed excitation sector of the Hilbert space. Calculating the imbalance in the eigenstates is analogous to calculating the mean imbalance in the trajectory of the system starting from these initial states. Similarly one can calculate the standard deviation in ${I}$ for a given trajectory using 
\begin{equation}
    \sigma_{I} = \sqrt{\frac{1}{\mathbb{N}}\sum_{i=1}^\mathbb{N}\left({I}_i - \expval{{I}} \right)^2},
    \label{eq:classical_STD}
\end{equation} 
where $\expval{{I}}$ is the mean value of ${I}$, and ${I}_i$ is the value of ${I}(t)$ at the $i^{\rm th}$ time step and $\mathbb{N}$ is the total number of time steps. In Fig.~\ref{fig:classical_mean_vs_rms} we plot $\sigma_{I}$ vs. $\expval{{I}}$ for the classical TCD with $N_{\rm p} = 5/2$ and we find {that remarkably it has the same scattered structure as the quantum case} [Fig.~\ref{fig:quantum mean_vs_std}] for both values of $\lambda$ in the two different regimes. {At first glance it might seem that the quantum mechanical standard deviation [Eq.~\eqref{eq:QM_STD}] and the classical standard deviation {[Eq.~\eqref{eq:classical_STD}]} are incompatible for comparison as the first one is the uncertainty in the state and the second one is the fluctuation in time around the mean value. However, one can envisage the quantum fluctuation in the state as the smearing of the observable around the mean value (which remains constant for an eigenstate), akin to classical fluctuation.}

% This is not the case, since one can think of the quantum fluctuation in the state to be the smearing of the observable around the mean value (which remains constant for an eigenstate) which is similar to the classical fluctuation.

\section{Open Tavis-Cummings dimer}
\label{sec:openTCD}

\begin{figure}[h]
    \centering
    \includegraphics[scale = 0.35]{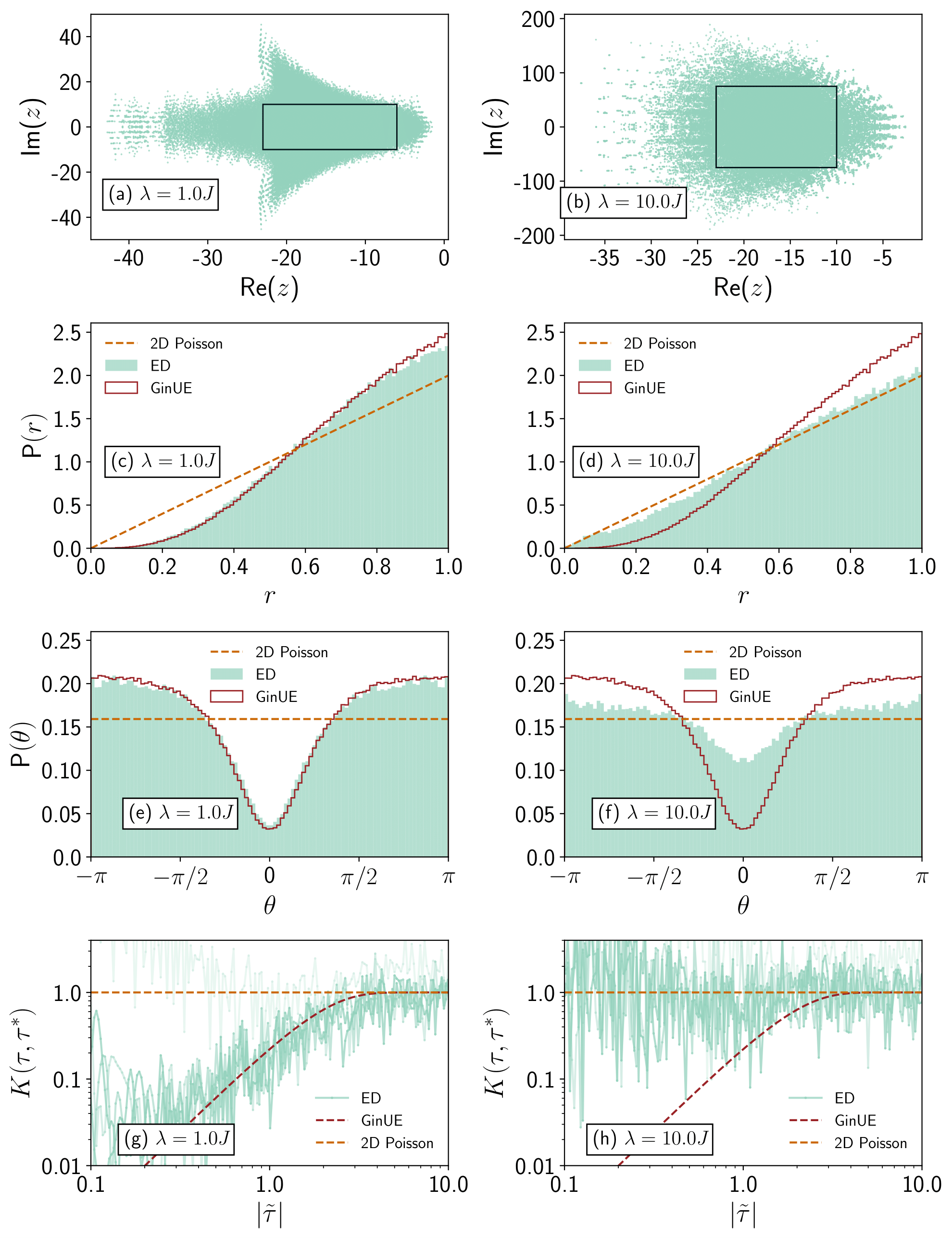}
    \caption{The eigenvalues for the open TCD Liouvillian, described in Eq.~\eqref{eq:lindbladME}, is plotted in the complex plane for (a) $\lambda = 1.0J$ and (b) $\lambda = 10.0J$. The rectangular box in (a) and (b) indicates the eigenvalues selected for the spectral analysis. The distribution of the absolute value of the CSR, $r$ is plotted for (c) $\lambda = 1.0J$ and (d) $\lambda = 10.0J$ for the open TCD discussed in Sec.~\ref{sec:openTCD}. In (e) and (f) the distribution of the argument of the CSR $\theta$ is plotted for $\lambda = 1.0J$ and $\lambda = 6.8J$ respectively. In (g) and (h) we have plotted the dissipative spectral form factor as a function of rescaled time $\tilde{\tau} = \tau/\tau_H$ for $\lambda = 2.0J$ and $\lambda = 10.0J$ respectively. For all the plots we have considered 6 realizations of the Liouvillian with $N_{\rm cut} = 3$, $S = 3$, and with a 10$\%$ disorder in the real part of the cavity frequency. The cavity dissipation rate $\kappa = 1.0J$.}
    \label{fig:CSR_openTC}
\end{figure}

In this section, we present results for the open TCD where we couple the cavity mode at each site with a Markovian bath. The effect of the bath is modeled using the Lindblad operators and the evolution of the system density matrix is given by the Liouvillian superoperator 
\begin{equation}
\dot{\rho} = \mathcal{L}\rho \, . 
\end{equation}
The superoperator 
$\mathcal{L}$ is defined as
\begin{equation}
    \mathcal{L}\star = -\text{i}\left[H_{\rm TCD},\star \right] + \sum_k \left[\mathcal{O}_k\star \mathcal{O}_k^\dagger - \frac{1}{2}\{\mathcal{O}^\dagger_k\mathcal{O}_k,\star \}\right],
    \label{eq:lindbladME}
\end{equation}
where $\mathcal{O}_k$ are the Lindblad jump operators. For our system, the jump operators are given by,
\begin{eqnarray}
    \mathcal{O}_1 &=& \sqrt{2 \kappa}a_L\\
    \mathcal{O}_2 &=& \sqrt{2 \kappa}a_R^{\dagger}\, ,
    \label{eq:jump_operators}
\end{eqnarray}
where $\kappa$ is the cavity dissipation rate. {Such a model corresponds to a gain-loss dynamics in the system.} To classify the system into a chaotic or regular system we investigate the spectral statistics of $\mathcal{L}$, which is non-hermitian. As mentioned before, the closed TCD system has a $U(1)$ symmetry where the total excitation number $\hat{N}$ is conserved. In the case of the open system, the $\mathcal{L}$ has a weak $U(1)$ symmetry $\mathcal{N}_-$ , such that $\left[\mathcal{L},\mathcal{N}_- \right] = 0$. The weak $U(1)$ symmetry is defined as
\begin{equation}
    \mathcal{N}_-\star = \left[\hat{N},\star\right],
    \label{eq:weak_U1_symm}
\end{equation}
A more detailed discussion on this weak symmetry {in open TCD} is mentioned in App.~\ref{app:weak_symm_of_liouv}. {It is important to note that the number of excitations is no longer conserved and we need to introduce a cut-off $N_{\rm cut}$ for the cavity modes.} We compute the spectrum of $\mathcal{L}$ using ED and compute the complex spacing ratio (CSR)
\begin{equation}
    \xi_n = \frac{z^{NN}_n - z_n}{z^{NNN}_n - z_n} = r_n e^{\text{i}\theta_n}
    \label{eq:CSR}
\end{equation}
where $z^{NN}_n$ and $z^{NNN}_n$ are the nearest and next-nearest neighbor of the complex eigenvalue $z_n$ of the Hamiltonian, and $r_n$ and $\theta_n$ are the absolute value and argument of $\xi_n$ respectively. The distance between two eigenvalues is measured by the absolute value of the difference between the two values. This diagnostic tool is independent of the distribution of the underlying spectrum and hence does not require any unfolding. Furthermore, as done in the closed hermitian case, we also compute {a generalization of the} spectral form factor for non-hermitian systems, known as the dissipative form factor (DSFF) $K(\tau,\tau^*)$ \cite{PhysRevLett.127.170602,ghosh2022spectral},
\begin{equation}
    K(\tau,\tau^*) = \expval{\sum_{m,n}^\mathcal{N} e^{\text{i}\vec{\tau}.\vec{z}_{mn}}},
    \label{eq:DSFF}
\end{equation}
where $\tau = t + \text{i}s$ is the generalized time variable and $\vec{z}_{mn} = \{\text{Re}(z_m) - \text{Re}(z_n),\text{Im}(z_m) - \text{Im}(z_n)\}$ is the difference between eigenvalues $z_m$ and $z_n$. The averaging is done over disorder realizations. For computing the DSFF efficiently it is convenient to write it in the following form,
\begin{equation}
    K(\tau,\tau^*) = \expval{\abs{\sum_n^\mathcal{N} e^{\text{i}\left(z_n \tau^* + z_n^* \tau \right)/2}}^2},
    \label{eq:DSFF_num}
\end{equation}
which has a single sum over the eigenvalues instead of two. 
%\sout{\mk{It is important to note that although the DSFF is defined for a complex spectrum, it is independent of the $\phi = arg(\tau)$. This is due to the uniform and rotationally symmetric distribution of the complex eigenvalues in the complex plane.}} 
Much like the hermitian case, the signature for chaos can be seen from the presence of a ramp in $K(\tau,\tau^*)$. { To avoid capturing any edge effects of the spectrum, we consider the inner eigenvalues for each realization while computing the {spectral statistics}. In Fig.~\ref{fig:CSR_openTC} (a-b) we indicate the eigenvalues selected with a rectangular box. We compute the CSR [Eq.~\ref{eq:CSR}] using these selected eigenvalues. In Fig.~\ref{fig:CSR_openTC} (c-f), we plot the distribution of $r$ and $\theta$ described in Eq.~\eqref{eq:CSR} for $\lambda = 1.0J$ (left column) and $\lambda = 10.0J$ (right column). In Fig.~\ref{fig:CSR_openTC} (g-h) we plot the DSFF as a function of the $\abs{\tau/\tau_H}$, where $\tau_H$ is the Heisenberg time, for both the chaotic and regular regimes. In all the diagnostic tools used, we find that for $\lambda = 1.0J$ the spectral statistics agree with that of non-Hermitian RMT, in particular, the GinUE symmetry class, and for $\lambda = 10.0J$ we find $2D$ Poisson statistics}. {Although there is a marked distinction between $\lambda = 1.0J$ and $\lambda = 10.0J$, we notice a slight deviation from the 2D Poisson statistics in the $\lambda = 10.0J$ case. This is indicative of the fact that the mixed nature of the open quantum system still has not yet disappeared at $\lambda = 10.0J$. It is noteworthy that such remnants of mixed behavior are also evident in the closed quantum system case for $\lambda = 10.0J$ [see for e.g. Fig.~\ref{fig:spectral_statistics} b and d].}

{Regarding the computation of the SFF in the Hermitian case, we recall that} we unfolded the spectrum to bring out the underlying universal features. {However, in the case of DSFF,} for a complex spectrum, there is no {well-established} method to unfold the spectrum. Thus, to capture the universal features we need to scale $\abs{\tau}$ by the Heisenberg time $\tau_H$. This Heisenberg time for our system can be computed numerically by fitting the ramp in the data {of the RMT regime} with the {fitting function} $f_K(\tau,m) = 1 - \text{e}^{-m\abs{\tau}^2}$ by varying $m$, and $\tau_H$ is given by $\tau_H = 1/2\sqrt{m}$ \cite{ghosh2022spectral}. This fitting function is inspired by the analytical form of the DSFF for GinUE matrices.

\section{Non-hermitian Tavis-Cummings Dimer}
\label{sec:NHTCD}

\begin{figure}[h]
    \centering
    \includegraphics[scale = 0.35]{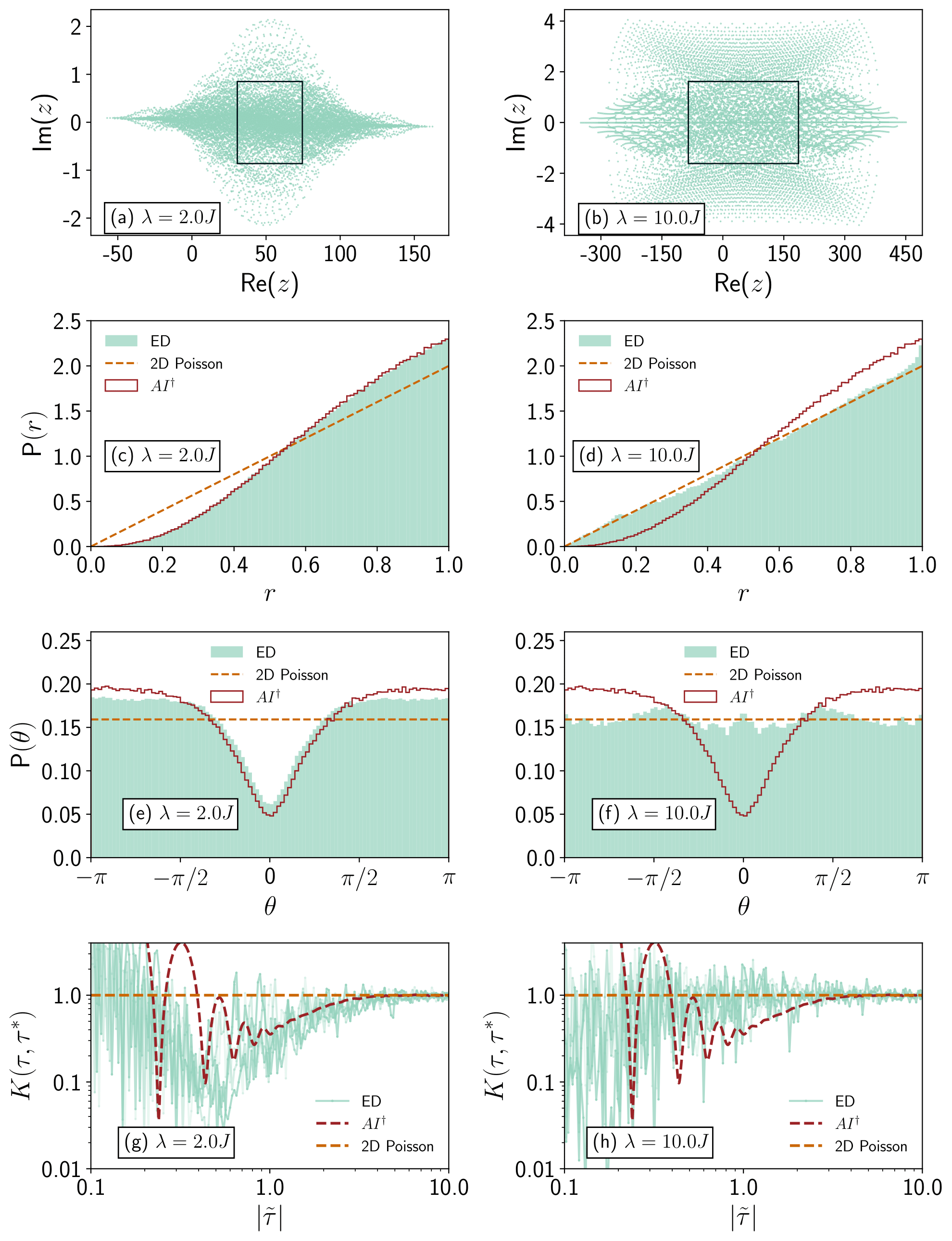}
    \caption{{The eigenvalues for the non-hermitian TCD Hamiltonian, described in Eq.~\eqref{eq:NH_TCD_hamil}, is plotted in the complex plane for (a) $\lambda = 2.0J$ and (b) $\lambda = 10.0J$. The rectangular box in (a) and (b) indicates the eigenvalues selected for the spectral analysis.} The distribution of the absolute value of the CSR, $r$ is plotted for (c) $\lambda = 2.0J$ and (d) $\lambda = 10.0J$ for a non-hermitian TCD {discussed in Sec.~\ref{sec:NHTCD}}. In (e) and (f) the distribution of the argument of the CSR $\theta$ is plotted for $\lambda = 2.0J$ and $\lambda = 10.0J$ respectively. In (g) and (h) we have plotted the dissipative spectral form factor as a function of rescaled time $\tilde{\tau}$ for $\lambda = 2.0J$ and $\lambda = 10.0J$ respectively. For all the plots we have considered 500 realizations of $H_{\rm TCDNH}$ {with $N_{\rm p} = 50$, $S = 20$ and} with a 10$\%$ disorder in the real part of the cavity frequency.  The imaginary part of the cavity frequency (non-hermiticity) is set to {$1.0J$ and $-1.0J$ for the left and right units respectively}. }
    \label{fig:CSR_nonherm}
\end{figure}

In this section, we show that the analysis done in this paper can also be extended to the non-hermitian {generalization of the TC} model where the Hamiltonian at each site is given by
\begin{equation}
    H_{\rm TCNH} =  (\omega_c + \text{i}\Delta) a^\dagger a + \omega_S S^z + \frac{\lambda}{\sqrt{S}}\left(a^\dagger S^- + a S^+ \right),
\end{equation}
and the dimer Hamiltonian is described by
\begin{equation}
    H_{\rm TCDNH}  = H_{\rm TCNH}^L + H_{\rm TCNH}^R + J\left(a^\dagger_L a_R + a_R^\dagger a_L\right).
    \label{eq:NH_TCD_hamil}
\end{equation}
% Such non-Hermitian models are not only physical but also have been realized experimentally \cite{lee2014heralded, lapp2019engineering}, where the non-hermiticity is introduced in the cavity frequency. 
Even in the case of non-Hermitian Hamiltonians, we consider terms that mimic the gain-loss mechanism.  To achieve this, we set $\Delta = \kappa$ for the left unit and $\Delta = -\kappa$ for the right unit. {Such gain-loss mechanisms in oscillators, realized as a non-Hermitian Hamiltonian, have been studied in Refs.~\onlinecite{119519,lange2020rotation}.} Since the Hamiltonian preserves the transposition symmetry, this model belongs to the $AI^\dagger$ symmetry class and hence the spectral statistics for the chaotic regime should conform with this symmetry class. {As was done in the case of the Liouvillian, to avoid capturing any edge effects of the spectrum, we consider the 5000 inner eigenvalues for each realization while computing the DSFF. In Fig.~\ref{fig:CSR_nonherm} (a-b) we indicate the eigenvalues selected with a rectangular box. We compute the CSR described in Eq.~\ref{eq:CSR} using these selected eigenvalues.} In Fig.~\ref{fig:CSR_nonherm} (c-f), we plot the distribution of $r$ and $\theta$ described in Eq.~\eqref{eq:CSR} for $\lambda = 2.0J$ (left column) and $\lambda = 10.0J$ (right column). In Fig.~\ref{fig:CSR_nonherm} (g-h) we plot the DSFF as a function of the $\abs{\tau/\tau_H}$, where $\tau_H$ is the Heisenberg time, for both the chaotic and regular regimes. In all the diagnostic tools used, we find that for $\lambda = 2.0J$ the spectral statistics agree with that of {non-Hermitian RMT, in particular,} the $AI^\dagger$ symmetry class, and for $\lambda = 10.0J$ we find $2D$ Poisson statistics.

\section{Conclusions and outlook}

{To conclude, using the Tavis-Cummings dimer as a platform,} we explore two facets of quantum {systems} which are extremely relevant, both from theoretical as well as experimental perspectives. Firstly, we do a detailed analysis of the relationship between the macroscopic phenomenon of self-trapping and the integrability of the system. We employ tools from RMT to show self-trapping in regions of integrability and delocalization in regions of chaos. We show that self-trapping in the dimer can be achieved by increasing the light-matter interaction strength $\lambda$. While doing so, the spectral statistics of the system agree with uncorrelated Poisson levels. When $\lambda$ is small, the system is delocalized and its energy levels become correlated and the spectral statistics agree with RMT. However, we report that the transition from GOE statistics to Poisson statistics is not abrupt, and the transition does not seem to become sharper on increasing the local Hilbert space dimension. This indicates the possible presence of a mixed behavior regime. 

In the second part of our investigation, we delve into the concept of quantum-classical correspondence, which is defined by the Berry-Tabor and the BGS conjectures. {The TCD used in this paper is precisely suited for such a study as it has a well-defined classical limit.} Specifically, we address the presence of mixed phase space in the classical model. Recent studies have explored mixed phase space in various models, including the three-site Bose-Hubbard model \cite{Nakerst:2022prc}, kicked top model \cite{wang2023power}, and the Dicke model \cite{wang2023mixed}. Our study contributes to the burgeoning interest in understanding mixed phase spaces within quantum systems and their classical limits. In particular, we take the classical limit of the TCD and show similar transitions from delocalized to self-trapped states as was shown in the exact quantum model. Furthermore, we compute the Lyapunov exponent for an ensemble of initial states uniformly chosen over the manifold of fixed energy and excitation. We show signatures of mixedness in the Lyapunov distribution {which is consistent with the signatures of mixedness seen in the quantum system based on RMT diagnostics.}

Furthermore, we also extend our analysis of the model to an open quantum system where we connect the two sites of the dimer to Markovian baths that mimic gain-loss dynamics. We compute the CSR and DSFF of the Liouvillian spectrum for the delocalized and the self-trapped regime and show agreement with the statistics of the GinUE symmetry class and 2D Poisson statistics respectively. Our analysis of open quantum systems shows the possible existence of mixed behavior. Finally, we consider a non-Hermitian gain-loss Hamiltonian, where the non-hermiticity is introduced by making the cavity frequency complex. We show that for smaller values of $\lambda$, the system agrees with the $AI^\dagger$ symmetry class, and for larger values of $\lambda$ the system shows 2D Poisson statistics.

Our findings make important contributions to the existing literature on macroscopic self-trapping, quantum chaos, quantum-classical correspondence, and mixed phase space. Although our analysis was numerically extensive, there are several challenges given the complexity of the model. For example, the classical TCD model has eight DOFs, which makes analysis using Poincare maps very difficult, thereby leading to significant challenges while probing the mixed behavior of the system. As a future direction, it would be interesting and numerically more accessible to consider simpler models with fewer DOFs to study the presence of mixed phase spaces and their signatures in the quantum model. Furthermore, studying a non-hermitian model with the non-hermiticity introduced in the interaction term \cite{PhysRevE.80.026213} would be interesting.

\section*{Acknowledgements}
We gratefully acknowledge David Huse for many fruitful discussions and important suggestions during the course of this work. We thank Mahaveer Prasad and Soumi Ghosh for useful discussions. We acknowledge support from the Department of Atomic Energy, Government of India, under Project No. RTI4001. MK thanks the VAJRA faculty scheme (No. VJR/2019/000079) from the Science and Engineering Research Board (SERB), Department of Science and Technology, Government of India. MK acknowledges the hospitality of the Department of Physics, Princeton University where a major part of the work took place.

\appendix

\section{Details of diagnostics for chaos}
In this section, we discuss the other diagnostics used to identify signatures of chaos in a quantum system. In App.~\ref{App:level_spacing} we discuss the level spacing statistics and in App.~\ref{app:SFF} we discuss the spectral form factor.
\subsection{Level spacing}
\label{App:level_spacing}
One of the fundamental distinctions between a non integrable and an integrable quantum system is the presence and absence of level repulsion within the energy spectrum respectively. Therefore we compute the level spacing $s_n = E_{n+1} - E_n$ where the real spectrum $\{E_n \}$ is arranged in ascending order. Before computing the distribution, we perform the unfolding of the spectrum. This process of unfolding involves  the following steps - 
\begin{itemize}
    \item[i.] Evaluate the cumulative distribution function (CDF) of the eigen spectrum, $\hat{F}(E) = \sum_n \Theta(E - E_n)$, where $\Theta$ is the Heaviside step function.
    \item[ii.] Fit $\hat{F}(E)$ to a polynomial function F(E).
    \item[iii.] Construct the unfolded spectrum \{$\tilde{E}_n$ \} using,
    \begin{equation}
        \tilde{E}_n = F(E_n).
    \end{equation}
\end{itemize}

\subsection{Spectral form factor}
\label{app:SFF}
Yet another diagnostic tool used to differentiate between integrable and non integrable Hermitian quantum systems is the spectral form factor (SFF) of the unfolded spectrum.  The SFF, $K(t,\mathcal{N})$, which is defined as the Fourier transform of the two-point density correlation function $\expval{\rho(E)\rho(E+\omega)}$ , is defined for an ensemble of $N$ eigenvalues as,
\begin{equation}
    K(t,\mathcal{N}) = \expval{\sum_{m,n = 1}^\mathcal{N} e^{i t (E_m - E_n)}},
    \label{SFF_eq}
\end{equation}
where $\expval{\dots}$ indicates the average over an ensemble of realisations of these $N$ eigen values. The SFF probes correlations in the spectrum on scales inversely proportional to $t$. The behavior of $K(t,N)$ can be broadly described for three different regions in $t$ \cite{haake1991quantum,PhysRevResearch.3.L012019, Mahaveer_Abhishodh} - 
\begin{itemize}
    \item[i.] For small values of $t << \tau_T$, $K(t,\mathcal{N})$ probes the spectrum on the scale of the bandwidth. This part is sensitive to the tail of the spectrum.
    \item[ii.] For $\tau_T < t < \tau_H$, $K(t,\mathcal{N})$ is dominated by the universal global correlations, if present.
    \item[iii.] For $t >> \tau_H$, $K(t,\mathcal{N})$ probes the spectrum at the level of the mean level spacing (which is set to one for the unfolded spectrum) where the levels are quantized. Without the presence of accidental degeneracies, this region is dominated by terms with $m = n$ in Eq.~\eqref{SFF_eq} and eventually saturates at $K \approx \mathcal{N}$.
\end{itemize}
$\tau_T \sim 1/\mathcal{N}$ is called the Thouless time and $\tau_H \sim 1$ is known as the Heisenberg time. 

\section{Classical limit}
\label{app:quantum_to_classical}
In this section, we demonstrate how to shift to the classical limit of the system. To construct the classical Hamiltonian, we follow the following steps - 
\begin{itemize}
    \item[i.] Write down all annihilation and creation operators in terms of the corresponding position $\left(x_{L/R}\right)$ and momentum $\left(p_{L/R}\right)$ operators and replace them with their real values.
    \item[ii.] Write down all the spin raising and lowering operators in terms of $S^x_{L/R}, S^y_{L/R}$ and $S^z_{L/R}$ which obeys $(S^{x}_{L/R})^2 + (S^{y}_{L/R})^2 + (S^{z}_{L/R})^2 = \
    \frac{N}{2}\left(\frac{N}{2}+1 \right)$.
    \item[iii.] Now we take the limit ${N} \to \infty$ and {transform} $\left(S^x_{L/R}, S^y_{L/R}, S^z_{L/R} \right) \mapsto \left(S^x_{L/R}, S^y_{L/R}, S^z_{L/R} \right)/{N}$. This leads us to a classical spin vector on a circle of radius 1/2.
    \item[iv.] Finally we also {transform} $\left(x_{L/R},p_{L/R}\right) \mapsto \left(x_{L/R},p_{L/R}\right)/\sqrt{{N}}$.
\end{itemize}
Following the above steps, we get the classical TC Hamiltonian,
\begin{equation}
    H_{\rm clTC} = \frac{1}{2}\left(p^2 + \omega_c x^2\right) + \omega_s S^z + \lambda\sqrt{\frac{2}{\omega_c}} \left(\omega_c x S^x - p S^y \right),
\end{equation}
and the dimer Hamiltonian is given by,
\begin{equation}
    H_{\rm clTCD} = H_{\rm clTC}^L + H_{\rm clTC}^R + \frac{J}{\omega_c}\left(\omega_c^2 x_Lx_R + p_Lp_R \right).
\end{equation}
The energy {transforms} as $E \mapsto E/{N}$. The resulting EOMs for the rescaled variables are obtained using Hamilton's equations 
\begin{equation}
     \dot{x} = \pdv{H}{p}, \ \dot{p} = -\pdv{H}{q}, \ \dot{\vec{S}} = \nabla_{\vec{S}}H_{\rm clTCD} \cross \Vec{S},
    \label{hamil_cl_eq}
\end{equation}
we get
\begin{eqnarray}
\label{eq:b4}
    \dot{x}_{L/R} &=& p_{L/R} - \lambda\sqrt{\frac{2}{\omega_c}}S^y_{L/R} + \frac{J}{\omega_c}p_{R/L},\\
    \dot{p}_{L/R} &=& -\omega_c^2 x_{L/R} - \lambda\sqrt{\frac{2}{\omega_c}}S^x_{L/R} - J\omega_c x_{R/L},\\
    \dot{S}^x_{L/R} &=& -\lambda\sqrt{\frac{2}{\omega_c}}p_{L/R}S^z_{L/R} - \omega_sS^y_{L/R},\\
    \dot{S}^y_{L/R} &=& -\lambda\sqrt{{2}{\omega_c}}x_{L/R}S^z_{L/R} +\omega_sS^x_{L/R}\\
    \dot{S}^z_{L/R} &=& \lambda\sqrt{\frac{2}{\omega_c}}\left(\omega_cx_{L/R}S^y_{L/R} + p_{L/R}S^x_{L/R} \right).
    \label{eq:classical_EOM}
\end{eqnarray}

\section{Sampling of classical initial states}
\label{app:sampling}

In this section, we write down in detail, how to sample initial states for classical computations. To ensure that we sample states from all over the manifold of a fixed excitation sector $N_{\rm p}$, we follow the following procedure - 
\begin{enumerate}
    \item[i.] Randomly select an excitation sector (left cavity mode, left spin mode, right cavity mode, and right spin mode) and subtract $\delta n \sim \mathcal{U}(-1,1)$ excitation from this sector, {where we recall that $\mathcal{U}(a,b)$ is a uniform distribution from a to b}.
    \item[ii.] Check the validity conditions for the new excitation value:
    \begin{enumerate}
        \item For the cavity mode: $\text{new excitation} \geq 0$.
        \item For the spin sector: $0 \leq \text{new excitation} \leq 1$.
    \end{enumerate}
    If the conditions are not met, return to step 1, {else proceed to step 3.}
    \item[iii.] Randomly choose another sector different from the previous step and add $\delta n$ excitation to this sector. Check the validity conditions again. If the conditions are not met, repeat this step else, accept the state.
    \item[iv.] Ignore the first 5000 acceptable states. Starting from the $5001st$ state, collect every 200th valid state obtained by repeating steps 1-3.
\end{enumerate}
To facilitate the addition and removal of excitation from the excitation sectors, we employ the following protocols:

\textit{Cavity sector:} The excitation of the cavity sector is represented by
\begin{equation}
    N_{\rm p}^{\rm cav} = \frac{\omega_c}{2} x^2 + \frac{p^2}{2 \omega_c}.
    \label{eq:cavity_excitation}
\end{equation}
Consequently, the quantities $\sqrt{\omega_c}x$ and ${p}/{\sqrt{\omega_c}}$ lie on a circle with a radius of $\sqrt{2 N_{\rm p}^{\rm cav}}$. To determine the new state, we randomly select $\theta \sim \mathcal{U}(0, 2\pi)$. The new values of $x$ and $p$ corresponding to the new excitation are given by:
\begin{eqnarray}
x &=& \sqrt{\frac{2 (N_{\rm p}^{\rm cav}\pm \delta n)}{\omega_c}} \cos \theta,\\
p &=& \sqrt{2 \omega_c (N_{\rm p}^{\rm cav}\pm \delta n)} \sin \theta.
\end{eqnarray}

\textit{Spin sector:} The excitation of the spin sector is described by 
\begin{equation}
    N_{\rm p}^{\rm spin} = S^z + 1/2.
    \label{eq:spin_excitation}
\end{equation}
The new value of $S^z$ is given by:
\begin{equation}
S^z = (N_{\rm p}^{\rm spin} \pm \delta n) - \frac{1}{2}.
\end{equation}
As the classical spin vector resides on a sphere with a radius of $1/2$, we must update $S^x$ and $S^y$ accordingly. For this purpose, we once again select $\theta \sim \mathcal{U}(0, 2\pi)$ randomly, and the new values are determined as follows:
\begin{eqnarray}
S^x &=& \sqrt{\frac{1}{4} - (S^z)^2} \cos \theta,\\
S^y &=& \sqrt{\frac{1}{4} - (S^z)^2} \sin \theta.
\end{eqnarray}

\section{Calculation of maximal Lyapunov exponent}
\label{app:lyap}
In this section, we discuss the algorithm used to compute the Lyapunov exponent {using method described in Refs.~\onlinecite{benettin1976kolmogorov,benettin1980lyapunov,benettin2}.} We start with two initial states (A and B) such that they differ by a phase space distance of $\delta_0$. We let them evolve with the $H_{\rm clTCD}$ [Eqs.~\eqref{eq:b4}-\eqref{eq:classical_EOM}] for a time $\tau$ and calculate the phase space distance $\delta_1 = \norm{\textbf{x}_B(\tau)-\textbf{x}_A(\tau)}$ between the two trajectories. We then reset trajectory B to the initial distance from trajectory A, keeping the direction fixed,
\begin{equation}
    \textbf{x}_B^{\rm reset} = \textbf{x}_A + \delta_0 \frac{\textbf{x}_B(\tau)-\textbf{x}_A(\tau)}{\delta_1}.
\end{equation}
This process is repeated $M$ times and the finite time Lyapunov exponent is given by - 
\begin{equation}
    \Lambda_M = \lim_{\delta_0 \to 0}\frac{1}{M\tau}\sum_{j = 1}^M\log\left(\frac{\delta_j}{\delta_0}\right).
\end{equation}
The maximal Lyapunov exponent $\Lambda$ is obtained by considering the limit $M\tau \to \infty$. It must be noted that the value of $\tau$ must be chosen such that it is much less than the time taken for the Lyapunov to saturate. In our computations, we set $\tau = 1$. 

An important point to note is that when we reset the trajectories, conserved quantities such as energy, total excitation, and spin length {may} no longer {be} conserved. This is an inherent issue with {this resetting} method of computing the Lyapunov {exponent}. Nevertheless, since trajectory B differs from trajectory A only by an absolute difference of $\delta_0$ after resetting, we can assume that the violation of the conserved quantities is small {(depends on $\delta_0$)}.  In our computations, we set $\delta_0 = 10^{-8}$.

\section{Weak symmetries of the Liouvillian}
\label{app:weak_symm_of_liouv}

In this section, we show that the weak $U(1)$ symmetry defined in Eq.~\eqref{eq:weak_U1_symm} is indeed a symmetry of the Liouvillian given in Eq.~\eqref{eq:lindbladME}. For this we have to show that 
\begin{equation}
    \left[\mathcal{L},\mathcal{N}_-\right] = 0.
\end{equation}
For simplicity, we will only consider the left cavity dissipation channel with the jump operator $\mathcal{O} = \sqrt{2\kappa}a_L$.
% The calculation for the full Lindbladian simply follows from the left-right symmetry of the whole system and the distributive property of the commutator.
We expand the action of the commutator on a generic state $\rho$ as follows - 
\begin{eqnarray}
    \left[\LL,\Nm\right](\rho) &=&  \LL\Nm(\rho) - \Nm \LL(\rho) \hspace{3cm}\nonumber\\
                                &=& \LO\left(\N\rho - \rho \N\right)+ \LJ\left(\N\rho - \rho \N \right)\nonumber\\
                                &&- \left[\N \LO(\rho) - \LO(\rho)\N\right] \nonumber\\
                                && - \left[\N\LJ(\rho) - \LJ(\rho)\N\right]\nonumber\\
                                &=& -\frac{1}{2}\Big[\OO^\dagger\OO\left(\N\rho - \rho \N\right) \nonumber\\
                                && \hspace{1cm} + \left(\N\rho - \rho \N\right)\OO^\dagger\OO\Big] \nonumber\\
                                && + \frac{1}{2}\Big[\N\left(\OO^\dagger\OO\rho + \rho \OO^\dagger\OO\right) \nonumber\\
                                && \hspace{1cm} + \left(\OO^\dagger\OO\rho + \rho \OO^\dagger\OO\right)\N\Big] \nonumber\\
                                && + \LJ\left(\N\rho - \rho \N \right) \nonumber \\
                                &&-\left[\N\LJ(\rho) - \LJ(\rho)\N\right]
\label{eq:weak_symm_proof_1}
\end{eqnarray}
where we have used the fact that $\left[H_{\rm TCDNH},\N\right] = 0$. Furthermore, using $\left[\OO^\dagger\OO,\N\right] = 0$, we can show that the first four terms in Eq.~\eqref{eq:weak_symm_proof_1} cancel out and we are left with
\begin{eqnarray}
    \left[\LL,\Nm\right](\rho) &=& \LJ\left(\N\rho - \rho \N \right)-\left[\N\LJ(\rho) - \LJ(\rho)\N\right] \nonumber\\
    &=& \OO\left(\N\rho - \rho \N \right)\OO^\dagger - \N\OO\rho\OO^\dagger + \OO\rho\OO^\dagger\N \nonumber \\
    &=& 0.
\end{eqnarray}
In the final step, we have used the relations $\left[\OO,\N\right] = \OO$ and $\left[\OO^\dagger,\N\right] = -\OO^\dagger$. {It can be shown that the above calculation is valid even when we consider $\mathcal{O} = \sqrt{2\kappa}a_R^\dagger$ with a few minor changes.} 

\bibliography{bibliography}
\end{document}